\documentclass[12pt]{article}

\usepackage{graphicx}
\usepackage{epsf}

\usepackage{amsmath,amssymb,float} 

\usepackage{pstricks}
\usepackage{color}
\usepackage{multirow}
\usepackage{cite}

\def\beq{\begin{equation}}
\def\eq{\end{equation}}
\def\eeq{\end{equation}}

\def\centeron#1#2{{\setbox0=\hbox{#1}\setbox1=\hbox{#2}\ifdim
\wd1>\wd0\kern.5\wd1\kern-.5\wd0\fi
\copy0\kern-.5\wd0\kern-.5\wd1\copy1\ifdim\wd0>\wd1
\kern.5\wd0\kern-.5\wd1\fi}}
\def\ltap{\;\centeron{\raise.35ex\hbox{$<$}}{\lower.65ex\hbox{$\sim$}}\;}
\def\gtap{\;\centeron{\raise.35ex\hbox{$>$}}{\lower.65ex\hbox{$\sim$}}\;}

\def\chii0{\chi_i^0}
\def\chij0{\chi_j^0}

\def\foursqr#1#2{{\vcenter{\vbox{
 \hrule height.#2pt
 \hbox{\vrule width.#2pt height#1pt \kern#1pt
 \vrule width.#2pt}
 \hrule height.#2pt
 \hrule height.#2pt
 \hbox{\vrule width.#2pt height#1pt \kern#1pt
 \vrule width.#2pt}
 \hrule height.#2pt
     \hrule height.#2pt
 \hbox{\vrule width.#2pt height#1pt \kern#1pt
 \vrule width.#2pt}
 \hrule height.#2pt
     \hrule height.#2pt
 \hbox{\vrule width.#2pt height#1pt \kern#1pt
 \vrule width.#2pt}
 \hrule height.#2pt}}}}
\def\psqr#1#2{{\vcenter{\vbox{\hrule height.#2pt
 \hbox{\vrule width.#2pt height#1pt \kern#1pt
 \vrule width.#2pt}
 \hrule height.#2pt \hrule height.#2pt
 \hbox{\vrule width.#2pt height#1pt \kern#1pt
 \vrule width.#2pt}
 \hrule height.#2pt}}}}
\def\sqr#1#2{{\vcenter{\vbox{\hrule height.#2pt
 \hbox{\vrule width.#2pt height#1pt \kern#1pt
 \vrule width.#2pt}
 \hrule height.#2pt}}}}

\def\figin{\epsfcheck\figin}\def\figins{\epsfcheck\figins}
\def\epsfcheck{\ifx\epsfbox\UnDeFiNeD
\message{(NO epsf.tex, FIGURES WILL BE IGNORED)}
\gdef\figin##1{\vskip2in}\gdef\figins##1{\hskip.5in}
\else\message{(FIGURES WILL BE INCLUDED)}%
\gdef\figin##1{##1}\gdef\figins##1{##1}\fi}
\def\DefWarn#1{}
\def\figinsert{\goodbreak\midinsert}
\def\ifig#1#2#3{\DefWarn#1\xdef#1{fig.~\the\figno}
\writedef{#1\leftbracket fig.\noexpand~\the\figno}%
\figinsert\figin{\centerline{#3}}\medskip\centerline{\vbox{\baselineskip12pt
\advance\hsize by -1truein\noindent\footnotefont{\bf
Fig.~\the\figno:\ } \it#2}}
\bigskip\endinsert\global\advance\figno by1}

\def\fig#1#2#3#4{\vskip 0.5cm \begingroup \midinsert \centerline{
\psfig{file=#1,width=#2}} \vskip 0.4cm
\global\advance\figno by 1
\centerline{\vbox{\baselineskip=12pt \noindent Figure \the\figno: #3}}
\endinsert \endgroup {\xdef#4{\the\figno}} }

\def\figcrop#1#2#3#4#5#6#7#8{\vskip 0.5cm \begingroup \midinsert \centerline{
\psfig{file=#1,width=#2,bbllx=#3,bblly=#4,bburx=#5,bbury=#6}} \vskip 0.4cm
\global\advance\figno by 1
\centerline{\vbox{\baselineskip=12pt \noindent Figure \the\figno: #7}}
\endinsert \endgroup {\xdef#8{\the\figno}} \vskip .5cm}

\def\figlabel#1{\xdef#1{\the\figno}}
\def\encadremath#1{\vbox{\hrule\hbox{\vrule\kern8pt\vbox{\kern8pt
\hbox{$\displaystyle #1$}\kern8pt}
\kern8pt\vrule}\hrule}}
\def\underarrow#1{\vbox{\ialign{##\crcr$\hfil\displaystyle
 {#1}\hfil$\crcr\noalign{\kern1pt\nointerlineskip}$\longrightarrow$\crcr}}}

\begin{document}

\begin{titlepage}

\begin{center}
\vspace*{-1cm}

\vskip 0.7in
{\LARGE \bf Higgs Boson Contributions to the } \\
\vspace{.15in}
{\LARGE \bf Electron Electric Dipole Moment } \\
\vspace{.15in}

\vskip 0.35in
{\large Daniel Egana-Ugrinovic$^1$ and}~\!~~
{\large Scott Thomas$^2$}

\vskip 0.2in
$^1$
{\em C. N. Yang Institute for Theoretical Physics, 
Stony Brook, NY 11794}
\\
\vskip 0.07in
$^2$
{\em New High Energy Theory Center, 
Rutgers University, 
Piscataway, NJ 08854}

\vskip 0.7in

\end{center}

\baselineskip=16pt


\noindent

The contributions of a second Higgs doublet 
to the electron electric dipole moment near the heavy
Higgs decoupling limit are determined 
within an effective field theory framework. 
In models that 
satisfy the Glashow-Weinberg condition, 
the leading contributions in this limit at effective dimension six
are shown to come from two-loop Barr-Zee diagrams 
that include only internal Standard Model 
gauge bosons, 
third-generation fermions and 
the Standard Model-like Higgs boson. 
Additional contributions from two-loop Barr-Zee 
diagrams that include heavy Higgs bosons are sub-leading 
and contribute only at effective dimension eight near the decoupling limit.  
This simplification implies that to leading order in this limit, 
contributions of a second Higgs doublet to the electron electric dipole moment 
can be couched entirely in terms 
of the ratio of Higgs doublet expectation values and a single universal phase appearing in the 
effective couplings of the Standard Model-like Higgs boson to fermions, 
without direct reference to the heavy Higgs boson masses or couplings.  
The bound on the electron electric dipole moment 
from the ACME II experiment 
constrains the phases of the couplings of the Standard Model-like Higgs boson to 
up-type quarks and leptons at the part per mil level 
in Type I and IV two Higgs doublet models.   
In Type II and III models these phases are constrained at the two parts per mil or better level 
except in a tiny sliver of parameter space with nearly equal Higgs doublet expectation values
where destructive interference among contributing diagrams happens to occur. 
In a more general phenomenological parameterization with individual effective phases in the couplings 
of the Standard Model-like Higgs boson to third generation fermions and the electron, 
the top quark and electron coupling phases are constrained at the part per mil level except in tiny slivers 
of parameter space, 
while the bottom quark and tau-lepton coupling phases are constrained only at the thirty percent level.


\end{titlepage}

\baselineskip=17pt

\newpage


\section{Introduction}

Discovery of the Higgs boson at the Large Hadron Collider (LHC) \cite{Aad:2012tfa,Chatrchyan:2012xdj}
has opened up a new experimental window
into the physics responsible for electroweak symmetry breaking.  
Cross section times branching ratio measurements of the various Higgs boson production and decay 
channels that  have been isolated in LHC data 
provide a wealth of information about 
the coupling strengths of the Higgs boson to individual third-generation fermions and 
massive vector bosons \cite{Aad:2015gba,Khachatryan:2014jba,combination,ATLAS:2018doi,Aaboud:2018urx,ATLAS:2018nkp,ATLAS:2018lur,CMS:2018lkl,Sirunyan:2018hoz,CMS:2018nqp}. 
Additional and complementary information on the Higgs sector may be extracted from the short distance effects of the Higgs boson in high-precision, low-energy experiments. 
In this work,
we study the effects of parity (P) and time reversal (T) violation in the Higgs sector on the electron electric dipole moment (eEDM). 
We focus on this particular observable, 
first because it has been recently probed with exquisite sensitivity in the Advanced Cold Molecule Electron EDM 
thorium monoxide experiment ACME II \cite{Andreev:2018ayy},
and second 
unlike the neutron electric dipole moment or Schiff moments of nuclei, 
it does not suffer large hadronic matrix element uncertainties. 

P and T violation arise in many extensions of the Standard Model,
and T violation in the Higgs sector is particularly important for theories of electroweak baryogenesis 
\cite{McLerran:1990zh,Turok:1990zg,Cohen:1991iu,Zhang:1994fb,Cline:1995dg,Bodeker:2004ws,Fromme:2006cm,Fromme:2006wx,Cline:2011mm,Shu:2013uua,Dorsch:2016nrg,Basler:2017uxn,Egana-Ugrinovic:2017jib,Bruggisser:2017lhc,deVries:2017ncy,Huang:2018aja,Bruggisser:2018mrt}.
The bulk of this work is dedicated to studying theories in which P and T violation can be parametrized by a non-minimal Higgs sector extended to include a second Higgs doublet \cite{Lee:1973iz,Lee:1974jb,Fayet:1974fj}.
We further specialize in Two Higgs Doublet Models (2HDM) with Glashow-Weinberg conditions (types I-IV) \cite{Glashow:1976nt},
since such theories are protected from strongly constrained flavor changing neutral currents \cite{Buras:2010mh}.
In an appendix,
we also discuss other parametrizations allowing for P and T violation in the individual Higgs Yukawa interactions with the electron, tau, top and bottom fermions,
which arise in a variety of popular models as the aligned 2HDM \cite{Pich:2009sp,Gori:2017qwg} and in theories of electroweak baryogenesis \cite{Zhang:1994fb,Bodeker:2004ws}.

In the 2HDM with Glashow-Weinberg conditions,
the current generation of Higgs boson cross section times branching ratio measurements requires proximity to the limit in which the couplings of the 125 \textrm{GeV} Higgs boson are aligned with the Standard Model expectations \cite{Craig:2013hca}.
One way in which alignment can be obtained is in the heavy Higgs decoupling limit.
In this limit,  
the additional Higgs bosons beyond the Standard Model-like Higgs boson are made heavy by taking combinations of the dimensionful relevant Higgs potential parameters large, holding the dimensionless marginal interactions and measured Higgs sector parameters fixed \cite{Haber:1989xc,Gunion:2002zf}.  
Near this limit the couplings of the Standard Model-like Higgs boson are nearly aligned with those of the Standard Model Higgs boson.

Near the decoupling limit, 
the ratio of electroweak condensate to the masses of the heavy Higgs bosons provides a small decoupling parameter in which to analyze the contributions to the eEDM. 
In the exact decoupling limit the contributions of the second doublet to the eEDM vanish, 
while near the decoupling limit the contributions to the eEDM may be organized as an expansion in the small decoupling parameter.
A systematic method to obtain this expansion is to use effective field theory (EFT).
In this approach,  
the heavy Higgs doublet is integrated out and the short-distance effects on the low energy theory are organized in an operator product expansion (OPE) controlled by the operator's effective decoupling dimension \cite{Egana-Ugrinovic:2015vgy}.
In this work, 
we obtain the contributions from the second doublet to the eEDM at leading order (effective-dimension six) in the OPE,
or equivalently, 
at leading order in the small decoupling parameter.
We do not study the regime in which the masses of the extra Higgs bosons are comparable or smaller than the scale of the electroweak condensate.
In this case,  
alignment may be achieved without decoupling, and a different analysis of the eEDM is needed \cite{Craig:2013hca,Carena:2013ooa,Haber:2018ltt,Dev:2014yca}. 
\footnote{Even if alignment is ensured, 
 moderate heavy Higgs decoupling is still required in the types II-III 2HDM for all values of $\tan\beta$ and in the types I-IV 2HDM at low $\tan\beta$, 
for consistency with flavor bounds \cite{Arbey:2017gmh}.}

Near the decoupling limit, 
the EFT approach allows us to identify considerable simplifications in the calculation of the eEDM within the types I-IV 2HDM.
The most important result is that at effective-dimension six, 
P and T violation in the eEDM (or any other low energy observable) may be completely described by only two parameters, 
in addition to measured Standard Model couplings and masses.
The first parameter is the  ratio of the two Higgs doublet expectation values, $\tan\beta$.
The second parameter is a single P- and T-violating effective phase,
which may be very simply expressed in terms of a unique combination of relevant and marginal couplings of the two Higgs doublet potential.
This unique phase appears exclusively in effective dimension-six cubic Yukawa interactions of the light Higgs doublet with fermions.
Other operators containing different P- and T-violating combinations of 2HDM parameters, 
including the unknown individual masses of the heavy Higgs states, 
are suppressed at effective dimension-six and only arise at effective-dimension eight.
In particular, 
we show that P and T violation from heavy-Higgs mediated four-fermion interactions, 
arises only at effective-dimension eight due to a GIM cancellation between the contributions of the two heavy neutral Higgs bosons to the operator's coefficients. 

Using our results and the latest ACME II eEDM experimental limits,
we severely constrain P and T violation in all types I-IV 2HDM. 
We find that in all the types of 2HDM near the decoupling limit, 
the P- and T-violating part of the Higgs coupling to any fermion is constrained to be approximately less than $10^{-3}$ times the corresponding Standard Model Yukawa,
with two exceptions.
First, 
large P- and T-violating Higgs-fermion interactions are still allowed in the types II-III 2HDM in a narrow region around $\tan\beta \simeq 1.3$,
where destructive interference between different contributions to the eEDM arises. 
Second, 
the Higgs coupling with bottom quarks may still have a significant P- and T violating part in the type IV 2HDM at large $\tan\beta$.

Finally, 
and going beyond the types I-IV 2HDM, 
in an appendix we also briefly discuss ACME II limits on theories that can be parametrized by P and T violation in the individual Higgs Yukawa interactions with the electron, tau, top and bottom fermions. 
We find that,
excepting regions where interference between the contributions to the eEDM from the different P- and T-violating Yukawa interactions arises,
P and T violation in the electron and top Yukawas is constrained to be less than $2.0 \times 10^{-3}$ and $1.3 \times 10^{-3}$ times the corresponding Standard Model Yukawa.
P and T violation in the bottom and tau Yukawas may still be considerable,
at the level of $\simeq 0.3$ times the corresponding Standard Model Yukawa.

This paper is organized as follows. 
In section \ref{sec:two} we present the Naturally Flavor Conserving 2HDM and define notation.
In section \ref{sec:three} we study P and T violation in the effective theory obtained by integrating out the heavy second Higgs doublet.
In section \ref{sec:four} we obtain the leading order contributions to the eEDM for the types I-IV 2HDM using the EFT description,
and we use the recent ACME II limits to set constraints on P and T violation on the types I-IV 2HDM. 
In section \ref{sec:five} we discuss the general conclusions of our work. 
Appendices \ref{app:one}-\ref{app:three} are dedicated to technicalities. 
Finally, 
in appendix \ref{app:four} we present limits on P and T violation in the individual Higgs Yukawa interactions with the electron, tau, top and bottom fermions for generic theories beyond the 2HDM.

\section{Naturally flavor conserving two-Higgs doublet Higgs sector}
\label{sec:two}
We consider a Higgs sector containing two doublets $\Phi_{a}$, $a=1,2$, with hypercharge  $+1$. 
The most general Lagrangian density at the renormalizable level for the two doublets and Standard Model fields is
\begin{equation}
D_\mu \Phi_a^\dagger D^\mu \Phi_a
  - V(\Phi_1, \Phi_2) 
   - 
   \bigg[ ~
     \lambda^u_{aij} ~ Q_i \Phi_a \bar{u}_j
   - \lambda^{d\dagger}_{aij} Q_i \Phi_a^c \bar{d}_j 
   - \lambda^{\ell\dagger}_{aij} L_i \Phi_a^c \bar{\ell}_j 
  +{\rm h.c.}~ \!  \bigg]
  \label{eq:2HDMactiongeneric}
  \quad ,
\end{equation}
where the most general renormalizable potential for the two doublets is
\begin{eqnarray}
V( \Phi_1 , \Phi_2) 
&=&
m_{1}^2 \Phi_1^\dagger \Phi_1+
m_{2}^2 \Phi_2^\dagger \Phi_2 -
\Big(m_{12}^2 \Phi_1^\dagger \Phi_2 +{\rm h.c.} \Big) + 
\frac{1}{2}\lambda_1 ( \Phi_1^\dagger  \Phi_1)^2
\nonumber 
\nonumber \\
&+&
\frac{1}{2}\lambda_2 ( \Phi_2^\dagger  \Phi_2)^2 +
\lambda_3 ( \Phi_1^\dagger  \Phi_1)(\Phi_2^\dagger  \Phi_2) +
\lambda_4 ( \Phi_1^\dagger  \Phi_2)(\Phi_2^\dagger  \Phi_1)
\nonumber
\nonumber 
\\
&+& 
\bigg[ ~
      \frac{1}{2}\lambda_5(\Phi_1^\dagger \Phi_2)^2
      +\lambda_6(\Phi_1^\dagger \Phi_1)(\Phi_1^\dagger \Phi_2)
      +\lambda_7(\Phi_2^\dagger \Phi_2)(\Phi_1^\dagger \Phi_2)+{\rm h.c.} 
      ~\bigg]
      \quad .
      \label{eq:2HDMpotentialgeneric}
\end{eqnarray}
Here we consider the case in which the 2HDM potential leads to the usual spontaneous gauge symmetry breaking pattern, 
$SU(2)_L\times U(1)_Y \rightarrow U(1)_{em}$.
The corresponding gauge-invariant condensates for the fields $\Phi_1, \Phi_2$ are defined as
\begin{equation}
{v_1^2 \over 2} \equiv \langle \Phi_1^\dagger \Phi_1 \rangle  ~~~,~~~ 
{v_2^2 \over 2} \equiv \langle \Phi_2^\dagger \Phi_2 \rangle ~~~,~~~
{v_{12}^2 \over 2} \equiv \langle \Phi_1^\dagger \Phi_2 \rangle  ~~~,~~~
v_1^2+v_2^2 \equiv v^2 ~~~,
\label{eq:defvev}
\end{equation}
where $v_1$ and $v_2$ are real,  
$v_{12}$ is in general complex, 
and  $v=246 \, \textrm{GeV}$. 
We also define the ratio of the vacuum expectation values $\tan\beta$ and the relative phase of the condensates $\xi$
\begin{equation}
\tan \beta \equiv {v_2 \over v_1} 
\quad \quad , 
\quad \quad 
{\rm Arg} \langle \Phi_1^\dagger \Phi_2 \rangle \equiv \xi_2-\xi_1 \equiv \xi
\quad .
\label{eq:deftanb}
\end{equation}
Generically, 
the two-Higgs doublet theory leads to tree-level flavor-changing neutral currents (FCNC's) mediated by the neutral Higgs bosons,
which are strongly constrained experimentally.
Such tree-level FCNC's are avoided in two Higgs doublet theories with Glashow-Weinberg conditions \cite{Glashow:1976nt},
by requiring that each fermion representation acquires mass from only one of the two Higgs doublets.
This condition ensures that in the fermion mass eigenbasis the Yukawa matrices coupling neutral Higgs states to fermions are diagonal,
and may be imposed by assigning different charges under a discrete symmetry to the two doublets.
\footnote{Discrete symmetries only ensure the absence of tree-level FCNC's at leading order in a chiral expansion. 
For a review of higher order effects we refer the reader to \cite{Buras:2010mh}.}
The discrete symmetry also enforces that in the Higgs potential two of the quartic couplings vanish, 
namely
\begin{equation}
\lambda_6=\lambda_7=0 \quad .
\end{equation}
Glashow-Weinberg conditions lead to four possibilities for the fermion Yukawas of the two doublets, 
referred to as the types I-IV two Higgs doublet theories. 
The four possibilities are:
\begin{itemize}
\item Type I: all fermions couple only to one doublet $\lambda_{1}^{u,d,\ell}=0$.
\item Type II:  up-type quarks couple to one doublet, down-type quarks and leptons couple to the other doublet, $\lambda_1^u=\lambda_2^d=\lambda_2^{\ell}=0$.
\item Type III: quarks couple to one doublet, leptons couple to the other doublet, $\lambda_1^u=\lambda_1^d=\lambda_2^{\ell}=0$.
\item Type IV: up-type quarks and leptons couple to one doublet, down type quarks to the other, $\lambda_1^u=\lambda_2^d=\lambda_1^{\ell}=0$.
\end{itemize}
In the four types I-IV 2HDM,  
the Yukawa matrices are completely specified by the fermion mass matrices,
the Higgs condensate $v=246 \,\textrm{GeV}$, the ratio of the vacuum expectation values $\tan\beta$
and the condensate phase $e^{i\xi}$, 
so they do not introduce additional independent parameters to the theory beyond the ones already contained in the 2HDM potential. 
Explicit expressions for the Yukawa matrices may be found in \cite{Egana-Ugrinovic:2015vgy}.

In this work, 
we study the contributions of the four types of 2HDM with Glashow-Weinberg conditions to the eEDM. 
Two Higgs doublet models with Glashow-Weinberg conditions have a unique PQ invariant P- and T-violating phase,
which may be measured in the eEDM.
This phase may be chosen to be the relative phase between the complex quartic coupling and the condensate phase $\textrm{Arg}(\lambda_5 e^{2i\xi})$.
For brevity, 
for the rest of this work we refer to P and T violation simply as T violation, 
with the understanding that breaking of both symmetries is intended when referring to the 2HDM P- and T-violating phase.

The leading contributions of the 2HDM to the eEDM come from two-loop Barr-Zee diagrams with neutral Higgs mass eigenstates in the loops \cite{Barr:1990vd,Leigh:1990kf,Gunion:1990ce,Chang:1990sf,Abe:2013qla}.
The neutral Higgs mass eigenstates are admixtures of the neutral components of the doublets $\Phi_{1,2}$,
so their masses and couplings are usually obtained in a unitary mixing approach,
diagonalizing the $3\times 3$ neutral scalar mass-squared matrix.
The Barr-Zee diagrams mediated by the Higgs scalars depend on a combination of $\tan \beta$,  
the three neutral Higgs masses 
and three mixing angles, 
which in turn depend on complicated non-linear combinations of 2HDM parameters, 
including the T-violating phase $\textrm{Arg}(\lambda_5 e^{2i\xi})$.
As a consequence, 
available studies of the eEDM within a unitary mixing description
are presented slicing the multi-dimensional 2HDM parameter space \cite{Weinberg:1990me,Ipek:2013iba,Shu:2013uua,Bian:2014zka,Inoue:2014nva},
keeping some of the parameters fixed.

An alternative approach to the unitary mixing description may be used near the decoupling limit. 
In this limit, the masses of the extra Higgs bosons $H,A,H^\pm$ lie all close to a heavy scale $m_{H}\approx m_{A} \approx m_{H^\pm}$,
which is much larger than the electroweak scale $m_H \gg m_h = 125 \, \textrm{GeV}$.
Integrating out the heavy Higgs bosons leads to an effective theory (EFT) organized in an operator product expansion.
In the next section, 
we follow this EFT approach and we identify the T-violating operators contributing to the eEDM,
coming from integrating out the heavy Higgs bosons of the 2HDM.
In this way, 
we will show that near the decoupling limit,
several simplifications in the analysis of T violation in the types I-IV 2HDM arise.

\section{Dimension-six T-violating effective operators}
\label{sec:three}
In the decoupling limit, 
the heavy Higgs states may be integrated out,
and the low energy theory contains a single Higgs doublet $H$.
The modifications to the Standard Model from integrating out the heavy doublet are organized in an operator product expansion controlled by an effective operator dimension that counts powers of suppression by the heavy Higgs mass scale \cite{Egana-Ugrinovic:2015vgy}.
The effective operator dimension $n_E$ is defined by 
\begin{equation}
n_E=4-n_{m_H^2} \quad ,
\end{equation}
where $n_{m_H^2}$ corresponds to the number of powers of the heavy scale $m_H^2$ in the operator's coefficient. 
In this work, we work up to effective dimension-six.
Effective dimension-six operators capture the leading order corrections to the Standard Model in an expansion in powers of the electroweak scale over the heavy Higgs mass scale.
The effective dimension-six low energy effective theory obtained from integrating out the heavy Higgs bosons at tree level is \cite{Egana-Ugrinovic:2015vgy}
\begin{eqnarray}
 \nonumber &&  \! D_\mu H^\dagger  D^\mu H -V(H) -~
 \\
 && \! \! \! 
 \bigg[
Q_i H \big( \lambda^u_{ij} 
+\eta^{u }_{ij} H^\dagger  H \big) 
\bar{u}_j 
-Q_i H^c \big(  \lambda^{d\dagger}_{ij}  
+\eta^{d\dagger}_{ij}H^\dagger H \big)
\bar{d}_j
-L_i H^c \big( \lambda^{\ell\dagger}_{ij}  
+\eta^{\ell\dagger}_{ij}H^\dagger H \big)
\bar{\ell}_j 
+\textrm{h.c.} \bigg] \nonumber \\
&&\!\!\!
\nonumber \\
&& \! \! \! 
+ ~
\Omega^{u u ~\!\! (0)}_{ijmn }(Q_i \bar{u}_j)(\bar{u}_m^\dagger Q_n^\dagger)
+\Omega^{dd~\!\! (0)}_{ijmn }(Q_i \bar{d}_j)(\bar{d}_m^\dagger Q_n^\dagger)
+\Omega^{\ell\ell ~\!\! (0)}_{ijmn }(L_i \bar{\ell}_j)(\bar{\ell}_m^\dagger L_n^\dagger) \nonumber \\
&& \! \! \! + ~
\left[ \Omega^{d\ell ~\!\! (0)}_{ijmn }(Q_i \bar{d}_j)(\bar{\ell}_m^\dagger L_n^\dagger)
+\Omega^{u d ~\!\! (2)}_{ijmn }(Q_i \bar{u}_j)(Q_m \bar{d}_n) 
+\Omega^{u \ell ~\!\! (2)}_{ijmn }(Q_i \bar{u}_j)(L_m \bar{\ell}_n)+\textrm{h.c.} \right] 
\quad .
 \label{eq:2HDMsummary}
\end{eqnarray}
The coefficients of the dimension-six operators in \eqref{eq:2HDMsummary} and the dimension-six Higgs potential $V(H)$ are given in  \cite{Egana-Ugrinovic:2015vgy} for all the types I-IV 2HDM.
In what follows we concentrate on studying T violation in the effective theory. 
The effective theory \eqref{eq:2HDMsummary} contains only two T-violating dimension-six operators:
four-fermion and cubic Higgs Yukawa operators.
We now discuss the effects of each operator.

We start with the four-fermion operators. 
In \cite{Egana-Ugrinovic:2015vgy} it was found that for all the types I-IV two Higgs doublet UV completions, 
the two Higgs doublet theory T-violating phase $\textrm{Arg}(\lambda_5 e^{2i\xi})$ does not lead to T violation in the four-fermion operators at effective dimension-six,
due to a GIM cancellation.
We now reproduce a proof of this feature by calculating the different contributions to the four-fermion operator coefficients from each one of the heavy Higgs bosons.
The dimension-six four-fermion operator coefficients depend on the heavy-Higgs Yukawas and the heavy Higgs masses.
To keep track of the PQ invariant T-violating phase we work with PQ invariant heavy-Higgs Yukawas.
They are given by
\begin{eqnarray}
\nonumber
\lambda_{Hij}^f&\simeq& 
e^{{-\frac{i}{4}} \textrm{Arg}(\lambda_5 e^{2i\xi})} ~ \frac{m^{f}_{ij} }{v } 
\,
\frac{(v^2-v_f^2)^{1/2}}{v_f}
 \,
\bigg[
1+{\cal O}\bigg({\lambda v^2 \over {m_H}^2}\bigg)
\bigg]
\quad \quad ,
\\
\nonumber
\lambda_{Aij}^f&\simeq& 
i
e^{{-\frac{i}{4}} \textrm{Arg}(\lambda_5 e^{2i\xi})}
~ \frac{m^{f}_{ij} }{v } 
\,
\frac{(v^2-v_f^2)^{1/2}}{v_f}
 \, 
\bigg[
1+{\cal O}\bigg({\lambda v^2 \over {m_H}^2}\bigg)
\bigg]
\quad \quad ,
\\
\nonumber
\lambda_{H^+ ij}^{f_d \bar{f_u}}
&\simeq& 
e^{{-\frac{i}{4}} \textrm{Arg}(\lambda_5 e^{2i\xi})} 
~ \frac{\sqrt{2}m^{f_u}_{ij} }{v }
\,
\frac{(v^2-v_f^2)^{1/2}}{v_f}
 \,
\bigg[
1+{\cal O}\bigg({\lambda v^2 \over {m_H}^2}\bigg)
\bigg]
\quad \quad ,
\\
\lambda_{H^+ij}^{f_u \bar{f_d}}
&\simeq& 
e^{{-\frac{i}{4}} \textrm{Arg}(\lambda_5 e^{2i\xi})} 
~ \frac{\sqrt{2}m^{f_d}_{ij} }{v }
\,
\frac{(v^2-v_f^2)^{1/2}}{v_f}
 \,
\bigg[
1+{\cal O}\bigg({\lambda v^2 \over {m_H}^2}\bigg)
\bigg]
\quad \quad ,
\label{eq:Yukawasheavy}
\end{eqnarray}
where $v_f$ is the vacuum expectation value of the Higgs doublet giving mass to the fermion $f=u,d,\ell$,
so $(v^2-v_f^2)^{1/2}/{v_f}$ is $\cot\beta$ or $\tan\beta$ depending on the type of 2HDM.
In Eq. \eqref{eq:Yukawasheavy} we neglect overall signs shared by the Yukawas of the two neutral heavy Higgses,
 and overall signs of the charged Higgs Yukawas,
which are not important for this discussion.
The higher order corrections to Eq. \eqref{eq:Yukawasheavy} come suppressed by powers of $\lambda v^2/ {m_H}^2$,
where for the rest of this work $\lambda$ represents a combination of marginal Higgs potential couplings.
These corrections are not relevant for obtaining the dimension-six four-fermion operator coefficients.
It is important to note that all the Yukawas share the same factor of the universal T-violating phase, $e^{{-\frac{i}{4}} \textrm{Arg}(\lambda_5 e^{2i\xi})}$.

We start by integrating out the charged Higgs. 
This generates dimension-six four-fermion operators with coefficients 
\begin{equation}
\Omega^{f f'} 
\sim
\frac{
\lambda_{H^+}^{f \bar{f} \,*}
\,
\lambda_{H^+}^{f' \bar{f}'}}
{m_{H^\pm}^2}
\quad ,
\end{equation}
in which the universal phase $e^{{-\frac{i}{4}} \textrm{Arg}(\lambda_5 e^{2i\xi})}$ cancels trivially.
Integrating out the heavy scalar Higgs $H$ leads to four-fermion operators with coefficients proportional to $\sim \lambda_{H}^f \, \lambda_{H}^{f'}$ and $\sim \lambda_{H}^f \, \lambda_{H}^{f'\,*}$.
While in the latter the 2HDM T-violating phase cancels out,
the former four-fermion operators have T-violating  coefficients of the form
\begin{equation}
\Omega^{f f'}
\sim
\frac{
 \lambda_{H}^f \, \lambda_{H}^{f'}}
{m_H^2}
\sim
e^{-\frac{i}{2} \textrm{Arg}(\lambda_5 e^{2i\xi})}
~
\frac{
m^f m^{f'}/v^2...}
{m_H^2}
\quad .
\label{eq:fourfermionTH}
\end{equation}
Finally, 
integrating out the heavy pseudoscalar Higgs $A$ also leads to T-violating four-fermion operators with coefficients similar to Eq. \eqref{eq:fourfermionTH}, 
but with relative minus signs
\begin{equation}
\Omega^{f f'}
\sim
\frac{
 \lambda_{A}^f \, \lambda_{A}^{f'}}
{m_A^2}
\sim
i^2
e^{-\frac{i}{2} \textrm{Arg}(\lambda_5 e^{2i\xi})}
~
\frac{
m^f m^{f'}/v^2...}
{m_A^2}
=
-
e^{-\frac{i}{2} \textrm{Arg}(\lambda_5 e^{2i\xi})}
~
\frac{
m^f m^{f'}/v^2...}
{m_A^2}
\quad .
\label{eq:fourfermionTA}
\end{equation}
At lowest order in the decoupling parameter $H$ and $A$ are degenerate,
$m_H \sim m_A$,
so at effective dimension-six the two T-violating four-fermion operator coefficients Eq. \eqref{eq:fourfermionTH} and \eqref{eq:fourfermionTA} cancel out via a GIM mechanism in the Higgs sector,
as advertised above.
Note that the CKM phase does appear in effective dimension-six four-fermion operators, 
since it is contained in the quark mass matrices,
but its effects in the eEDM are strongly suppressed by the Jarlskog invariant  \cite{Jarlskog:1985ht}, 
and are neglected in what follows.

\begin{table}[h!]
\begin{center}
$
{\def\arraystretch{2}\tabcolsep=0pt
\begin{array}{|cc|c|c|c|c|}
\hline
&
&  		
\text{Type I} 
& 
\text{Type II} 
& 
\text{Type III} 
&
\text{Type IV} 
\\ 
\hline
\eta^u
&
\multirow{3}{*}{
$
=
~
\begin{array}{c}
\frac{1}{\sqrt{2} m_H^2} \sin 2\beta   \Big[\lambda_1 \cos^2 \beta-\lambda_2 \sin^2 {\beta} ~ -
\\
\lambda_{345}\cos 2\beta+i\textrm{Im}(\lambda_5 e^{2i\xi})\Big]
~~
\times
\end{array}
$}
&  	
 \cot\beta 
&
\cot\beta 
&  
 \cot\beta 
&
 \cot\beta
\\
\eta^d
&
&
 \cot\beta 
&
- \tan\beta 
&
 \cot\beta 
&
- \tan\beta 
\\
\eta^\ell
&
&
\cot\beta 
&
- \tan\beta 
&
-\tan\beta 
&
 \cot\beta 
\\
\hline
\end{array}}
$
\caption{
Operator coefficients for the effective dimension-six cubic Yukawa operators for the types I-IV 2HDM UV completions leading to the effective theory \eqref{eq:2HDMsummary}. 
The operator coefficients are given by $\eta_{ij}^f=\eta^f m^f_{ij}/v$, $f=u,d,\ell$,
and we use the definition $\lambda_{345}\equiv\lambda_3+\lambda_4+\lambda_5$.
}
\label{tab:Coefficients2}
\end{center}
\end{table}

The only remaining T-violating dimension-six operators are the cubic Higgs Yukawas,
coupling three insertions of the Higgs doublet field with Standard Model fermions.
The coefficients of the cubic Yukawa operators are given in table \ref{tab:Coefficients2}.
The relative phase between the fermion mass matrix $m^f_{ij}$ and the cubic Yukawa spurion $\eta^f_{ij}$ leads to T violation in the Higgs-fermion interactions.
We define the Higgs boson Yukawa couplings to up-type quarks, down-type quarks and leptons in the Lagrangian according to the conventions in \cite{Egana-Ugrinovic:2015vgy},
\begin{equation}
-h (\lambda^u_{hij} \, u_i \bar{u}_j + \lambda^{d \dagger}_{hij}\, d_i \bar{d}_j + \lambda^{\ell \dagger}_{hij} \, \ell_i \bar{\ell}_j + \textrm{h.c.})
\quad ,
\label{eq:yukawadef}
\end{equation}
where the Higgs boson Yukawas are given by
\begin{equation}
 \lambda^f_{h ij}
 =
 \frac{m^f_{ij}}{v}~\! 
 \bigg[
 1
  +
 2\bigg(\frac{\eta^f }{2\sqrt{2}}\bigg) v^2
 +
 \mathcal{O}
 \bigg(
 \frac{ \eta^f \lambda^2 v^4}
 {m_H^2},
  \frac{\lambda^2 v^4}
 {m_H^4}
 \bigg)
 \bigg]
 \quad .
 \end{equation}
and the cubic Yukawa coefficients $\eta^f$ for each fermion are given in table \ref{tab:Coefficients2}.
Committing to the fermion mass eigenbasis $m_{ij}^{f}=\delta_{ij} m_{i}^{f}$ with real mass eigenvalues $m_{i}^{f}$, the Higgs boson Yukawas may be written as
\begin{eqnarray}
 \lambda^{f}_{h  ij}
  \! \! \! &=& \! \! \!
\delta_{ij}~\!\frac{m_{i}^{f}}{v}
\, (\, 1+\kappa_{f} \, )
\quad ,
\label{eq:effyukdef}
\end{eqnarray}
where $\kappa_{f}$ are complex coupling modifiers given in table \ref{tab:Yukawas} for each type of fermion.
The imaginary part of the complex coupling modifiers  $\kappa_{f}$ represents T violation in the Higgs boson Yukawas.
By inspecting table \ref{tab:Yukawas}, 
we see that all T violation in the Higgs-fermion interactions for the types I through IV 2HDM, 
may be couched in terms of the ratio of the Higgs doublet expectation values $\tan\beta$, 
and a single effective phase $\delta_{h f \bar{f}}$ defined as
\begin{eqnarray}
\sin \delta_{h f \bar{f}}
&\equiv&
-
  \frac{1}{2}
\sin (2\beta)
\,
\textrm{Im}(\lambda_5 e^{2i\xi})
\frac{v^2}{m_H^2} \, 
\quad .
\label{eq:dim6phase}
\end{eqnarray}
Note that as expected,
the effective phase $\delta_{h f \bar{f}}$ vanishes in the exact decoupling limit,
and depends on the unique PQ invariant T-violating phase of the two Higgs doublet theory $\textrm{Arg} (\lambda_5 e^{2i\xi})$.
The effective dimension-six T-violating phase $\delta_{h f \bar{f}}$ may be expressed entirely in terms of Higgs potential couplings, 
as shown in appendix \ref{app:three}. 
The result is
\begin{eqnarray}
\sin \delta_{h f \bar{f}}
&=&
\frac{1}{2} \sin 2\beta |\lambda_5|
\sin \textrm{Arg}\big( m_{12}^4 \lambda_5^*\big)
\frac{v^2}
{
\,m_H^2
}
\bigg[
1+
{\cal O}\bigg(
\,
\frac{\lambda v^2}{m_H^2}
\,
\bigg)
\bigg]
\nonumber \\
\nonumber \\
&=&
\frac{\tan\beta}{1+\tan^2\beta}
 \,
|\lambda_5|
\sin \textrm{Arg}\big( m_{12}^4 \lambda_5^*\big)
\frac{v^2}
{
\,m_H^2
}
\bigg[
1+
{\cal O}\bigg(
\,
\frac{\lambda v^2}{m_H^2}
\,
\bigg)
\bigg]
\quad ,
\label{eq:explicitphase}
\end{eqnarray}
where in the last line we simply replaced $\sin 2\beta=2\tan\beta/(1+\tan^2\beta)$. 

The cubic Higgs operator  also leads to short-distance interactions between two and three Higgs bosons and two Standard Model fermions,
which also violate T due to the same effective phase $\delta_{h f \bar{f}}$.
These interactions are not relevant for the calculation of the eEDM at two-loops and are not discussed any further in this work. 
Summarizing, 
we find that all T violation beyond the CKM phase at dimension-six in the effective theory,
for the types I-IV two Higgs doublet UV completions,
is entirely due to the single effective dimension-six phase Eq. \eqref{eq:explicitphase} contained exclusively in the Higgs-fermion interactions.
This leads to a considerable simplification of the analysis of the eEDM within the 2HDM,
discussed in the next section.

We conclude this section by briefly commenting on the Higgs-gauge boson and gauge boson-fermion couplings, 
which are relevant for the calculation of the eEDM.
In the effective theory \eqref{eq:2HDMsummary} there are no operators with Higgs doublets and gauge bosons,
so we immediately see that the Higgs boson couplings to gauge bosons are left unmodified with respect to their Standard Model values at effective dimension-six,
\begin{equation}
g_{h VV}=\frac{2m_V^2}{v} \Bigg[1+{\cal O}\bigg({\lambda^2 v^4 \over {m_H}^4}\bigg) \Bigg]
\quad .
\label{eq:varphi1VVmixing}
\end{equation}
Finally, note that the heavy Higgs bosons do not have trilinear couplings to two gauge bosons at leading order in the decoupling parameter,
since near the decoupling limit they reside in the Higgs doublet that does not participate in electroweak symmetry breaking,
\begin{equation}
g_{{(H,A,H^\pm)} VV}={\cal O}\bigg({\lambda^2 v m_V^2 \over {m_H}^2}\bigg)
\quad .
\label{eq:varphi1VVmixing2}
\end{equation}
Due to the absence of heavy Higgs-gauge boson trilinear couplings, 
the effective theory \eqref{eq:2HDMsummary} does not contain short distance dimension-six interactions with two gauge bosons and two fermions. 

\begin{table}[ht]
\begin{center}
$
{\def\arraystretch{2}\tabcolsep=0pt
\begin{array}{|ccc|c|c|c|c|}
\hline
&
&
&  		
\text{Type I} 
& 
\text{Type II} 
& 
\text{Type III} 
&
\text{Type IV} 
\\ 
\hline
\textrm{Re} \, \kappa_{u}
&
\multirow{3}{*}{$ = $}
&
\multirow{3}{*}{$
~
\begin{array}{c}
 \frac{1}{2} 
\sin{2\beta}
(
\lambda_1 \cos^2{\beta}-\lambda_2 \sin^2{\beta}
\\
-\lambda_{345} \cos{2\beta})
\frac{v^2}{m_H^2}~\!
~~ \times
\end{array}
$}
&
 \cot~\!\beta 
&
 \cot~\!\beta 
&
  \cot~\!\beta 
&
  \cot~\!\beta 
\\
\textrm{Re} \, \kappa_{d}
&
&
&
\cot~\!\beta 
&
-\tan~\!\beta 
&
 \cot~\!\beta 
&
 - \tan~\!\beta 
\\
\textrm{Re} \, \kappa_{\ell}
&
&
&
\cot~\!\beta 
&
-\tan~\!\beta 
&
 -\tan~\!\beta 
&
  \cot~\!\beta 
\\
\hline
\textrm{Im} \, \kappa_{u}
&
\multirow{3}{*}{$=$}
&
\multirow{3}{*}{$
\underbrace{
  \frac{1}{2}
s_{2\beta}\,
\textrm{Im}(\lambda_5 e^{2i\xi})
\frac{v^2}{m_H^2} \, 
}_{ \equiv ~ - \, \sin \delta_{h f \bar{f}} ~\text{cf. Eq. \eqref{eq:dim6phase}}}
~~ \times ~~ $}
&
\cot \beta
&
\cot \beta
&
\cot \beta
&
\cot \beta
\\
\textrm{Im} \, \kappa_{d}
&
&
&
\cot \beta
&
-\tan\beta
&
\cot \beta
&
-\tan \beta
\\
\textrm{Im} \, \kappa_{\ell}
&
&
&
\cot \beta
&
-\tan\beta
&
-\tan\beta
&
\cot \beta
\\
\hline
\end{array}}
$
\caption{Higgs Yukawa coupling modifiers for the up-type quarks, down-type quarks and leptons at effective dimension-six for the types I-IV 2HDM.
The coupling modifiers are defined in Eq. \eqref{eq:effyukdef}.
In the table, $\lambda_{345}\equiv\lambda_3+\lambda_4+\lambda_5$, 
and we omit terms of order $\mathcal{O}(v^4/m_H^4)$,
which represent corrections of effective dimension-eight.
Note that in the 2HDM with Glashow-Weinberg conditions the coupling modifiers are family-universal, 
\textit{i.e.},
they are the same for the three fermion generations.}
\label{tab:Yukawas}
\end{center}
\end{table}

\section{Bounds on Higgs boson contributions to the eEDM}
\label{sec:four}

\begin{figure}[ht]
\begin{center}
\vspace*{.2in}
\includegraphics[width=10.2cm]{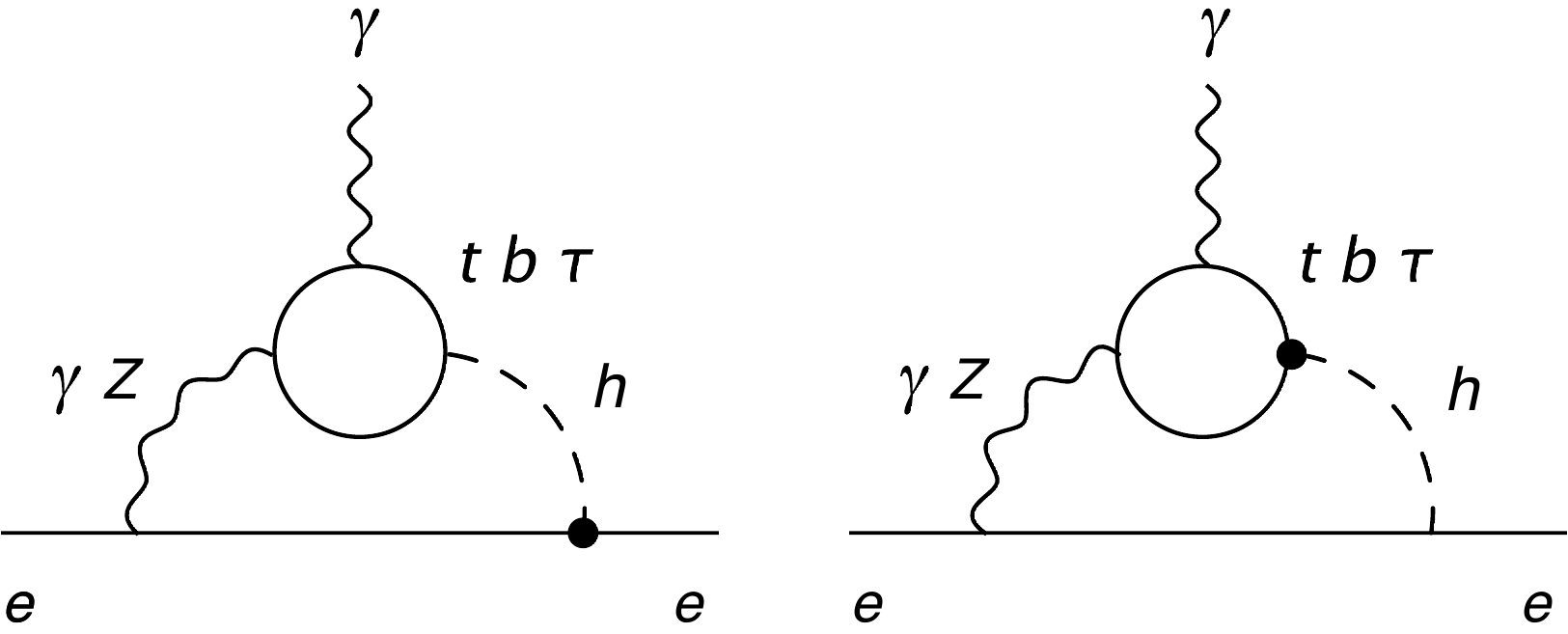}  
\caption{The leading two-loop contributions to the electron 
electric dipole moment that involve the Standard Model-like Higgs boson
and third-generation fermions 
in the low energy effective 
theory of 2HDMs in 
the heavy Higgs decoupling limit.
The dots represent parity and time-reversal violating Higgs-fermion interactions.
In the 2HDM, 
these interactions arise from  effective dimension-six operators generated by integrating out heavy Higgs bosons at tree-level.
The diagrams with a $Z$-boson propagator are suppressed by two orders of magnitude with respect to the diagrams with a photon propagator \cite{Barr:1990vd}, 
and are neglected.
}
\label{fig:BarrZee_D6_3rdgen}
\end{center}
\end{figure}

\begin{figure}[ht]
\begin{center}
\vspace*{.2in}
\includegraphics[width=10.2cm]{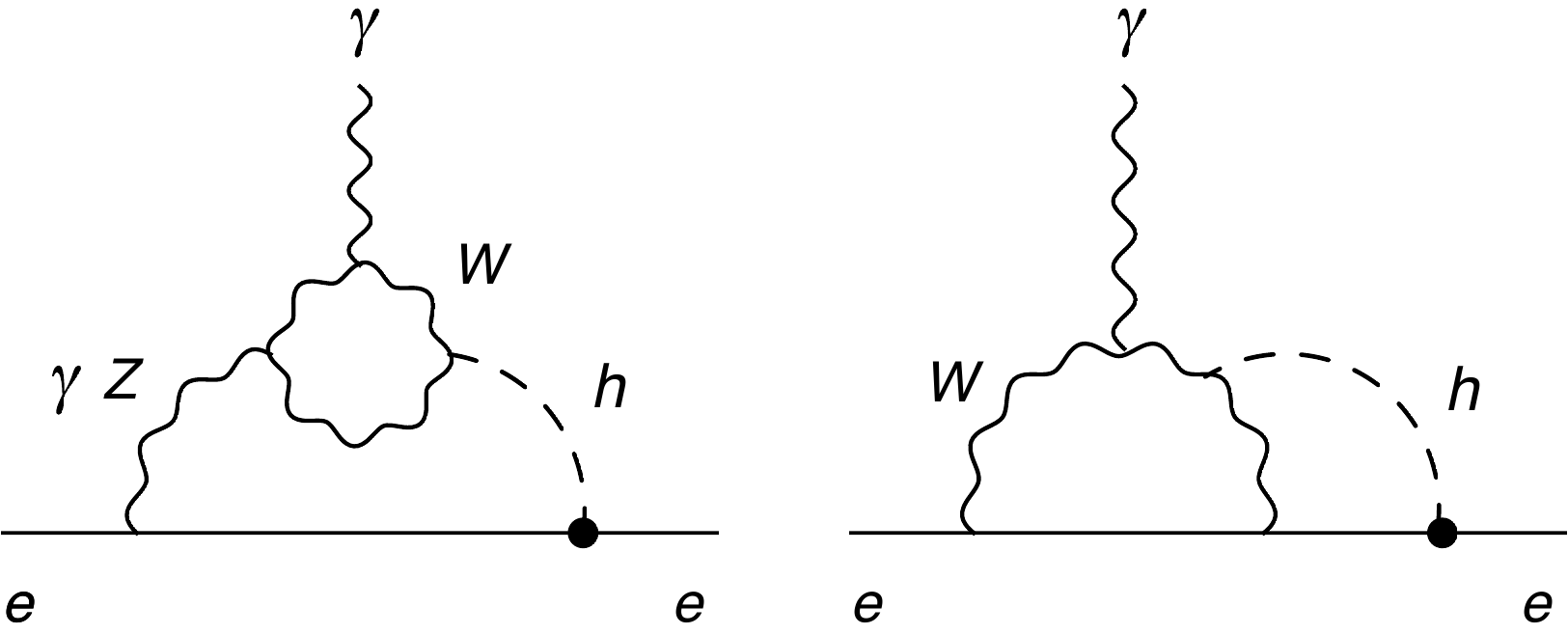}  
\caption{Some of the leading two-loop contributions to the electron 
electric dipole moment that involve the Standard Model-like Higgs boson
and $W$-boson 
in the low energy effective 
theory of 2HDMs in 
the heavy Higgs decoupling limit.
The dots represent parity and time-reversal violating Higgs-fermion interactions.
In the 2HDM, 
these interactions arise from  effective dimension-six operators generated by integrating out heavy Higgs bosons at tree-level.
Additional contributions that involve the Standard Model-like Higgs boson
and the $W$-boson as well as the $Z$-boson are not shown. 
The diagrams with a $Z$-boson propagator are suppressed by an order of magnitude with respect to the diagrams with a photon propagator and are neglected \cite{Leigh:1990kf}.
}
\label{fig:BarrZee_D6}
\end{center}
\end{figure}

\begin{figure}[ht]
\begin{center}
\vspace*{.2in}
\includegraphics[width=10.2cm]{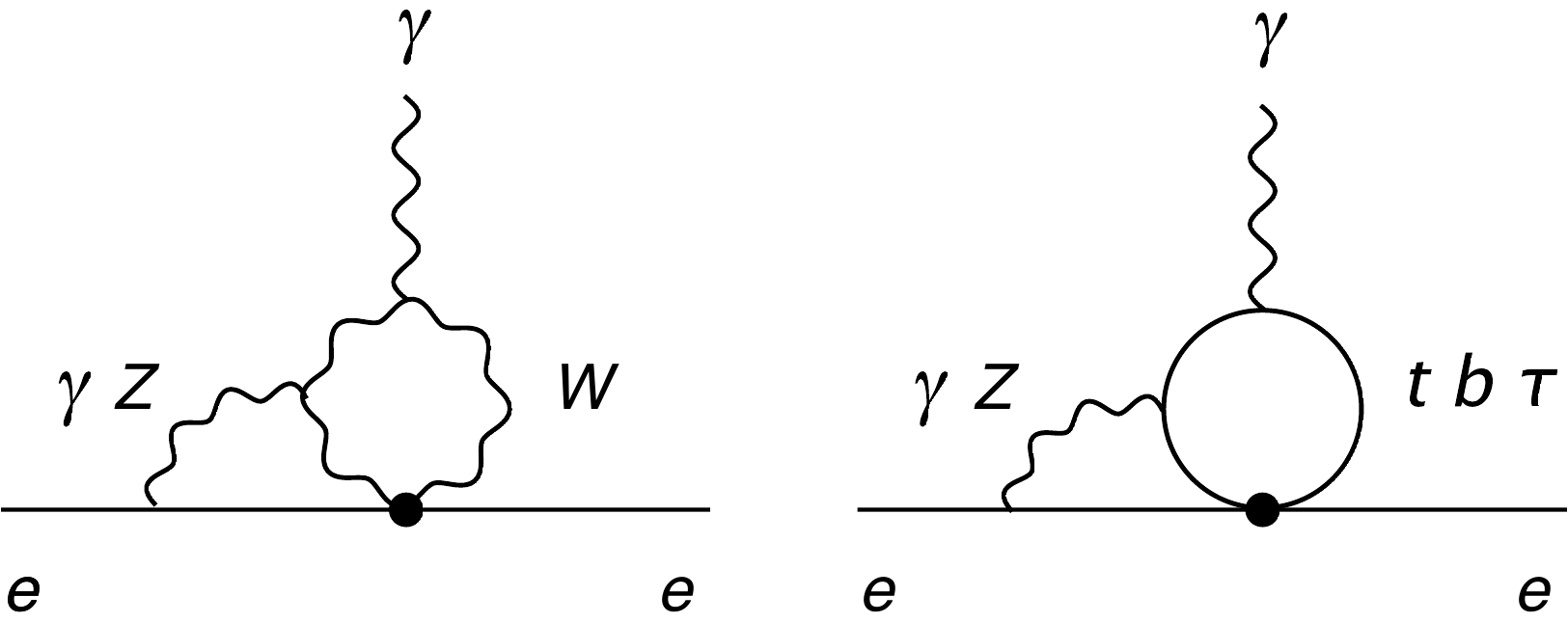}  
\caption{Some of the sub-leading two-loop contributions to the electron 
electric dipole moment that involve the $W$-boson or third-generation fermions 
in the low energy effective 
theory of 2HDMs in 
the heavy Higgs decoupling limit.
The dots represent parity and time-reversal violating 
operators of effective dimension-eight 
generated by integrating out heavy Higgs bosons at tree-level.  
Additional sub-leading two loop contributions that involve 
the $W$-boson as well as the $Z$-boson and charged Higgs boson are not shown. 
}
\label{fig:BarrZee_D8_nodot}
\end{center}
\end{figure}

\begin{figure}[ht]
\begin{center}
\includegraphics[width=16cm]{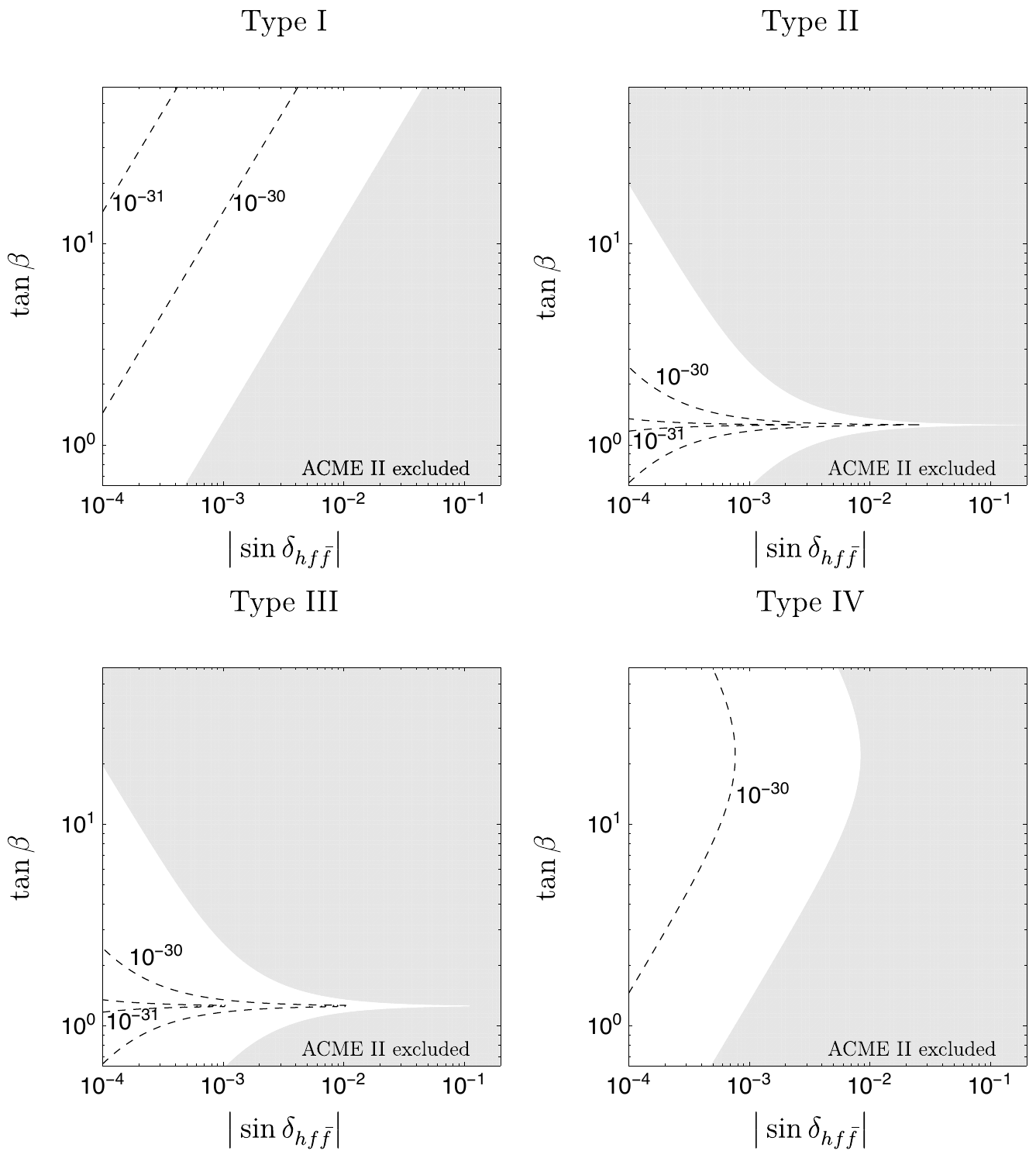}
\caption{The leading contributions to the 
magnitude of the electron electric dipole moment in units of $e$ cm 
near the heavy Higgs decoupling limit of the four types of 2HDMs
that satisfy the Glashow-Weinberg condition 
as a function of the ratio of the Higgs doublet expectation values, $\tan \beta$, 
and the single effective-dimension-six parity and time reversal violating phase that appears in 
Yukawa couplings of the Standard Model-like Higgs boson to fermions,
$\sin \delta_{h  f \bar{f} }$. 
The shaded region is excluded at $90\% \, \textrm{CL}$ by the current bound on the electron electric dipole moment of 
$|d_e| < 1.1 \times 10^{-29}$ $e$ cm from the ACME II experiment \cite{Andreev:2018ayy}. 
}
\label{fig:plots:one}
\end{center}
\end{figure}

\begin{figure}[ht]
\begin{center}
\vspace*{.2in}
\includegraphics[width=15.3cm]{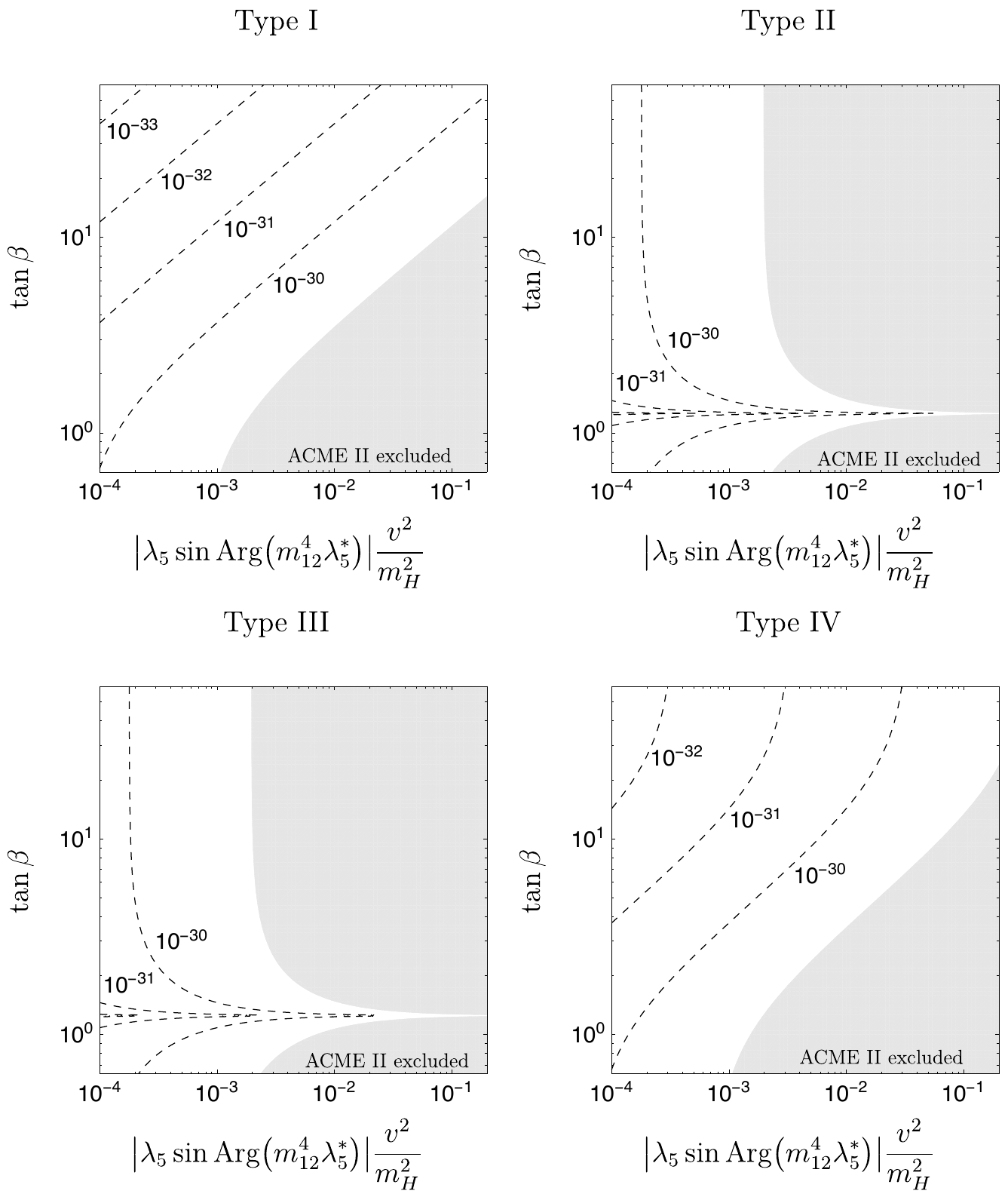}
\caption{The leading contributions to the 
magnitude of the electron electric dipole moment in units of $e$ cm 
near the heavy Higgs decoupling limit of the four types of 2HDMs
that satisfy the Glashow-Weinberg condition 
as a function of the ratio of the Higgs doublet expectation values, $\tan \beta$, 
and the Glashow-Weinberg basis Higgs potential parameters 
$|{\lambda_5}
\sin \textrm{Arg}( m_{12}^4 \lambda_5^*)|  \, {v^2 / m_H^2}$.
The shaded region is excluded at $90\% \, \textrm{CL}$ by the current bound on the electron electric dipole moment of 
$|d_e| < 1.1 \times 10^{-29}$ $e$ cm from the ACME II experiment \cite{Andreev:2018ayy}. 
}
\label{fig:plots:two}
\end{center}
\end{figure}

\begin{figure}[ht]
\begin{center}
\vspace*{.2in}
\includegraphics[width=16cm]{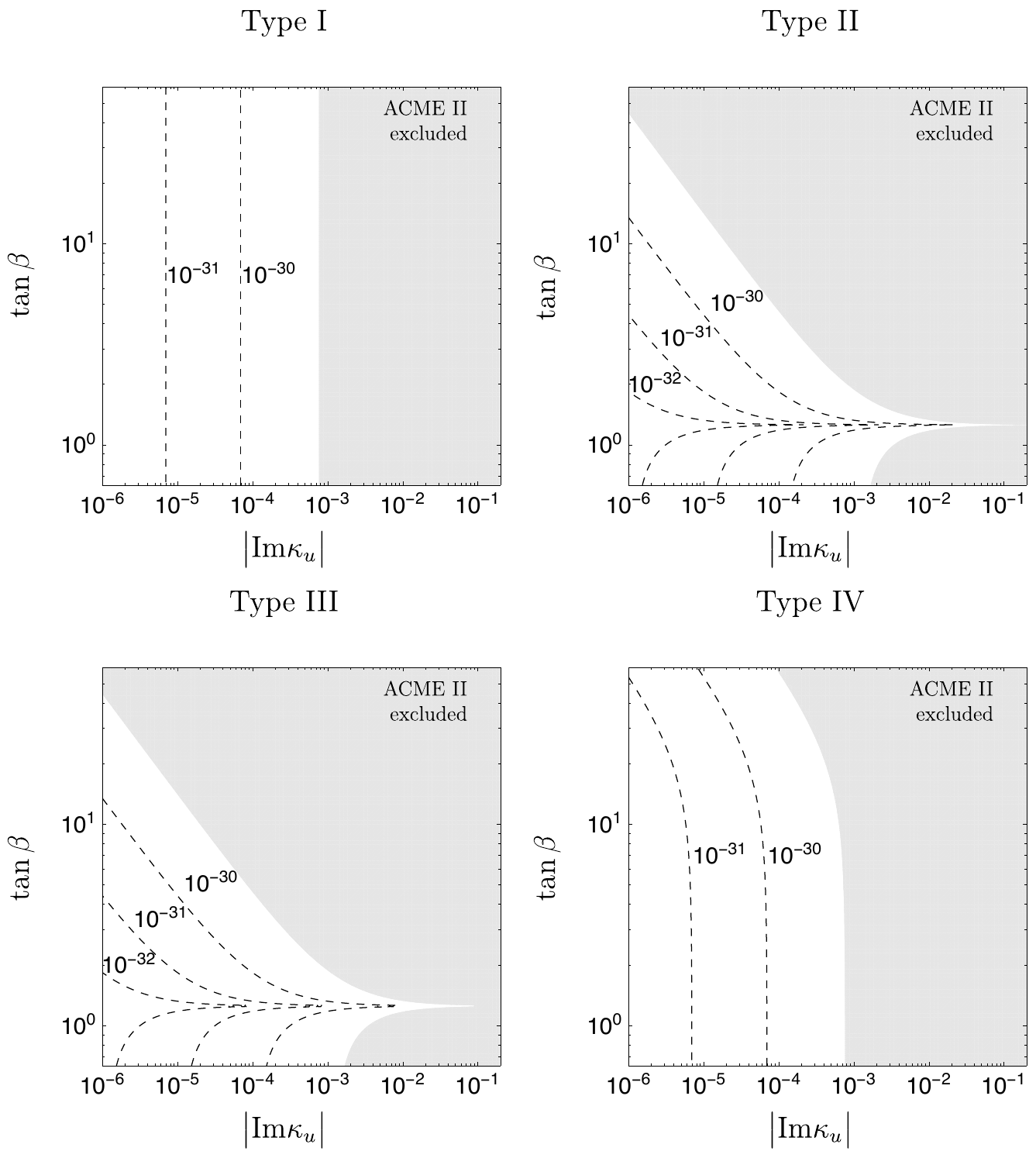}
\caption{The leading contributions to the 
magnitude of the electron electric dipole moment in units of $e$ cm 
near the heavy Higgs decoupling limit of the four types of 2HDMs
that satisfy the Glashow-Weinberg condition 
as a function of the ratio of the Higgs doublet expectation values, $\tan \beta$, 
and the imaginary part of the multiplicative modification of the Yukawa coupling 
of the Standard Model-like Higgs boson to the up-type quarks, ${\rm Im} \, \kappa_u$. 
The shaded region is excluded at $90\% \, \textrm{CL}$ by the current bound on the electron electric dipole moment of 
$|d_e| < 1.1 \times 10^{-29}$ $e$ cm from the ACME II experiment \cite{Andreev:2018ayy}.
}
\label{fig:plots:three}
\end{center}
\end{figure}

\begin{figure}[ht]
\begin{center}
\vspace*{.2in}
\includegraphics[width=16cm]{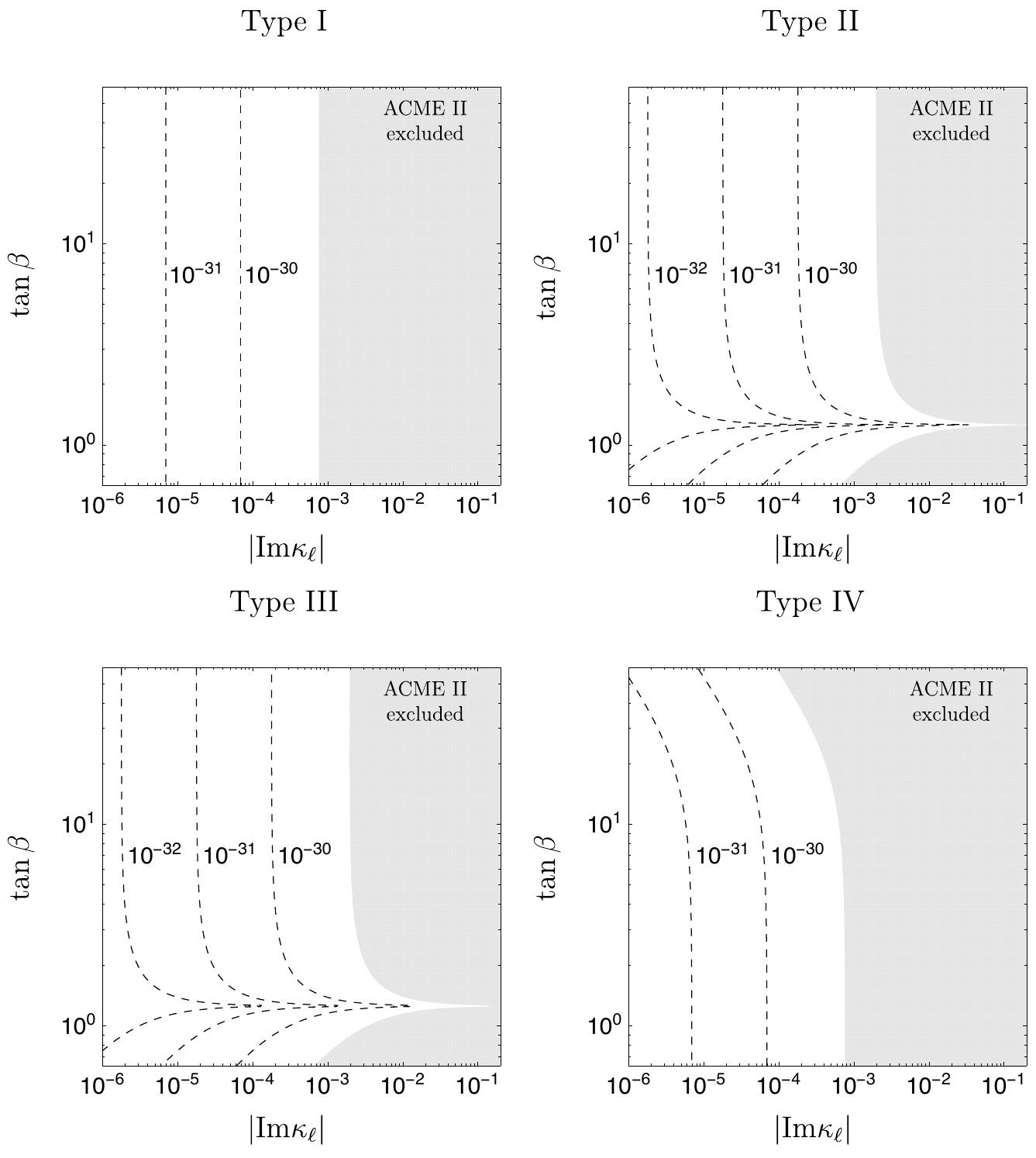}
\caption{The leading contributions to the 
magnitude of the electron electric dipole moment in units of $e$ cm 
near the heavy Higgs decoupling limit of the four types of 2HDMs
that satisfy the Glashow-Weinberg condition 
as a function of the ratio of the Higgs doublet expectation values, $\tan \beta$, 
and the imaginary part of the multiplicative modification of the Yukawa coupling 
of the Standard Model-like Higgs boson to leptons, ${\rm Im} \, \kappa_\ell$. 
The shaded region is excluded at $90\% \, \textrm{CL}$ by the current bound on the electron electric dipole moment of 
$|d_e| < 1.1 \times 10^{-29}$ $e$ cm from the ACME II experiment \cite{Andreev:2018ayy}. 
}
\label{fig:plots:four}
\end{center}
\end{figure}

\begin{figure}[ht]
\begin{center}
\vspace*{.2in}
\includegraphics[width=16cm]{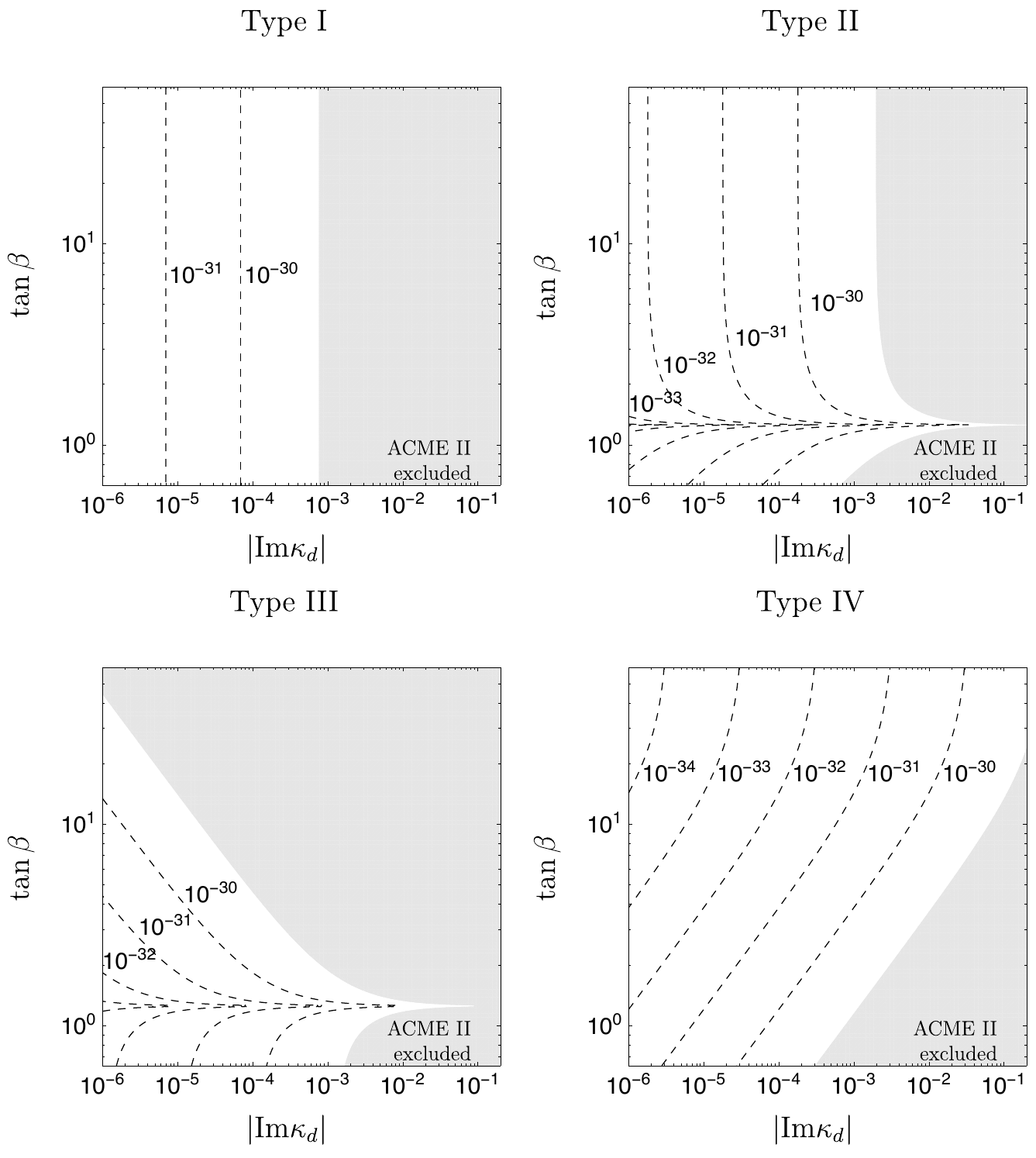}
\caption{The leading contributions to the 
magnitude of the electron electric dipole moment in units of $e$ cm 
near the heavy Higgs decoupling limit of the four types of 2HDMs
that satisfy the Glashow-Weinberg condition 
as a function of the ratio of the Higgs doublet expectation values, $\tan \beta$, 
and the imaginary part of the complex modifier of the Yukawa coupling 
of the Standard Model-like Higgs boson to down-type quarks, 
${\rm Im} \, \kappa_d$. 
The shaded region is excluded at $90\% \, \textrm{CL}$ by the current bound on the electron electric dipole moment of 
$|d_e| < 1.1 \times 10^{-29}$ $e$ cm from the ACME II experiment \cite{Andreev:2018ayy}. 
}
\label{fig:plots:five}
\end{center}
\end{figure}

P and T violation in the 2HDM leads to an eEDM at two loops via Barr-Zee diagrams \cite{Barr:1990vd,Leigh:1990kf,Gunion:1990ce,Abe:2013qla}.
In this section we constraint T violation in the types I-IV 2HDM using the latest $90\% \, \textrm{CL}$ ACME II experimental limit on the eEDM, 
given by \cite{Andreev:2018ayy}
\begin{equation}
d_e  < 1.1 \times 10^{-29} \, \, \textrm{e cm} \quad .
\label{eq:experimentalbound}
\end{equation}
As discussed in the previous section,
the only source of T violation in the types I-IV 2HDM in the effective theory up to dimension-six, 
is a universal T-violating phase in the Higgs Yukawa couplings.
This simplifies the determination of the eEDM, 
since T violation in the Higgs Yukawas leads only to two relevant types of Barr-Zee diagrams.
First, 
there are diagrams with third-generation quarks or leptons in the loops
with a T-violating Higgs coupling insertion on one of the fermion lines,
illustrated in  figure \ref{fig:BarrZee_D6_3rdgen}.
These diagrams have been calculated in \cite{Barr:1990vd,Jung:2013hka,Bian:2014zka}. \footnote{Our Yukawa coupling conventions for down type fermions differ from the ones in \cite{Barr:1990vd,Jung:2013hka,Bian:2014zka} by hermitian conjugation.}
Second, 
there are diagrams with $W$-bosons or ghosts in the loops illustrated in  figure \ref{fig:BarrZee_D6}, 
with a T-violating Higgs coupling insertion on the electron line,
which we obtain from \cite{Leigh:1990kf}.
The results in \cite{Leigh:1990kf} are presented in a unitary mixing description,
so to make use of them we derive the relation between the unitary mixing description and our effective theory notation in appendix  \ref{app:two}.
All the dimension-six gauge and fermion couplings of the Higgs needed for the calculation of the different diagrams are given in section \ref{sec:three}.

All the rest of the Barr-Zee diagrams contributing to the eEDM arise either at effective dimension-eight, 
or are numerically suppressed so they are neglected in what follows.
In particular and as discussed in the previous section,
diagrams with T-violating four-fermion interactions as in  figure  \ref{fig:BarrZee_D8_nodot} (right) are GIM suppressed so they arise only at dimension-eight.
Diagrams with interactions between two W bosons and two fermions as in  figure  \ref{fig:BarrZee_D8_nodot} (left), 
are of dimension-eight, cf. Eq. \eqref{eq:varphi1VVmixing2}.
Diagrams with charged Higgs bosons replacing the $W$-boson in  figure \ref{fig:BarrZee_D6} require insertions of neutral Higgs Yukawa T violation,
so they also arise at dimension-eight or higher.
Diagrams with $Z$-bosons are numerically small since they are suppressed by factors of $1-4\sin^2\theta_W $.
They lead at to an $\mathcal{O}(10\%)$ correction of our EDM results  \cite{Barr:1990vd,Leigh:1990kf},
so we neglect them.
Finally and for reference, 
the eEDM in the Standard Model arises first at four-loops, 
and the corresponding estimate is $d_e \leq 10^{-38}\, \, \textrm{e cm}$ \cite{Pospelov:1991zt}.

In summary, 
we obtain that the eEDM in terms of the T-violating Higgs interactions at effective dimension-six is given by
\begin{eqnarray}
\frac{d_e}{e} 
& \! \! \! \simeq \! \! \! &
 \frac{2\sqrt{2} \, \alpha \, G_F m_e}{(4\pi)^3} \bigg[ \, 
 -
\Big(
 - 
 \frac{16}{3}
 f(m_t^2/m_h^2)
 - \frac{4}{3}
 f(m_b^2/m_h^2)
 -4 \, 
f(m_\tau^2/m_h^2)
 \nonumber \\ 
 &&
\hspace*{2cm}  +~~
2 
~
\bigg[ 
~\! 
5g(m_W^2/m_h^2)+3 f(m_W^2/m_h^2) 
+
\frac{3}{4} \, 
\Big[
\,
g(m_W^2/m_h^2)+h(m_W^2/m_h^2)
\,
\Big]
~\! 
\nonumber \\
&&
\hspace*{2cm}-~~
\frac{g(m_W^2/m_h^2)-f(m_W^2/m_h^2)}{2 \, m_W^2/m_h^2}
\, \bigg]
+
\frac{1}{2\sin^2 \theta_W }
\,
D(m_W^2/m_h^2)
\,
\Big)
~
 \textrm{Im}\, \kappa_\ell
 \nonumber \\
 &&
\hspace*{2cm} - ~~
 \Big( \, 
 \frac{16}{3}
 g(m_t^2/m_h^2)
 \,
 \textrm{Im}\, \kappa_u
 - \frac{4}{3}
 g(m_b^2/m_h^2)
 \,
 \textrm{Im}\, \kappa_d
 - 4
 g(m_\tau^2/m_h^2)
 \,
  \textrm{Im}\, \kappa_\ell
 \,
 \Big)
 ~  
 ~
 \nonumber
 \\
 &&
\hspace*{2cm}  +~~
 \mathcal{O}
 \bigg(
 \frac{ \eta^f \lambda^2 v^4}
 {m_H^2},
  \frac{\lambda^2 v^4}
 {m_H^4}
 \bigg)
  \bigg]
 \label{edmformula}
\end{eqnarray}
where $\alpha$ is the fine structure constant, 
$G_F = 1/\sqrt{2} v^2$, $m_e=0.51 \, \textrm{MeV}$,
and the rest of the numerical parameters
are known Standard Model particle masses. 
The functions $f(z), g(z)$ and $D(z)$ and a summary of the Standard Model parameters used in this work are given in appendix \ref{app:one},
and the T-violating imaginary Higgs Yukawa modifiers $\textrm{Im}\, \kappa_{u,d,\ell}$ for the types I-IV 2HDM are given in table \ref{tab:Yukawas}.
Inserting the expressions in table \ref{tab:Yukawas} in the eEDM Eq. \eqref{edmformula} and evaluating the numerical prefactors we obtain,
\begin{eqnarray}
\nonumber
&\textrm{Type I:} & \!\!
d_e \simeq \, 1.43 \times 10^{-26}  \cot\beta \sin \delta_{h f \bar{f}}
\bigg[1+{\cal O}\bigg(\frac{\lambda^2 v^2}{{m_H}^2}\bigg) \bigg]
~\, \textrm{e cm}
\quad ,
\\ \nonumber
&\textrm{Type II:} & \!\!
d_e \simeq\big(8.77 \times 10^{-27} \cot\beta - 5.54\times 10^{-27} \tan\beta\big)  \sin \delta_{h f \bar{f}}
\bigg[1+{\cal O}\bigg(\frac{\lambda^2 v^2}{{m_H}^2}\bigg) \bigg]
~\, \textrm{e cm}
\quad ,
\\
&\textrm{Type III:} & \!\!
d_e \simeq\big(8.74 \times 10^{-27} \cot\beta - 5.57\times 10^{-27} \tan\beta\big)  \sin \delta_{h f \bar{f}}
\bigg[1+{\cal O}\bigg(\frac{\lambda^2 v^2}{{m_H}^2}\bigg) \bigg]
~\, \textrm{e cm}
\quad ,
\nonumber
\\
\nonumber
&\textrm{Type IV:} & \!\!
d_e \simeq\big(1.43 \times 10^{-26} \cot\beta + 2.96\times 10^{-29} \tan\beta\big)  \sin \delta_{h f \bar{f}}
\bigg[1+{\cal O}\bigg(\frac{\lambda^2 v^2}{{m_H}^2}\bigg) \bigg]
~\, \textrm{e cm}
\quad .
\\
\label{eq:finalEDM}
\end{eqnarray}
Expressions \eqref{eq:finalEDM} are the final result of this work. 
The eEDM for the types I-IV 2HDM up to effective dimension-six, 
can be expressed uniquely in terms of the universal T-violating dimension-six phase $ \delta_{h f \bar{f}}$ and the ratio of the Higgs vacuum expectation values $\tan\beta$. 

The electron electric dipole moment for the types I-IV 2HDM as a function of the dimension-six phase $\delta_{h f \bar{f}}$ and $\tan\beta$ is presented in  figure \ref{fig:plots:one}.
In the figure, 
we also show in gray the $90\% \, \textrm{CL}$ ACME II limits on the eEDM.
For the type I 2HDM at $\tan\beta = 1$,
the phase is constrained by the eEDM limit to be at the per mille level or below.
The limits are weaker at large values of $\tan\beta$
due to the corresponding suppression of the T-violating pieces of the Higgs Yukawas (see table  \ref{tab:Yukawas}).
For the types II and III 2HDM the phase is constrained to be below the per mille level for most values of $\tan\beta$, 
except in the region near $\tan\beta \simeq 1.3$, 
where a cancellation between different diagrams contributing to the eEDM occurs. 
In the types II and III 2HDM, the constraints are stronger at large values of $\tan\beta$ due to the corresponding enhancement of the T-violating lepton Higgs Yukawas.
The constraints for the type IV 2HDM are similar to the constraints for type I, 
but are stronger at large values of $\tan\beta$ due to a small contribution to the EDM coming from the T-violating coupling of the Higgs to bottom quarks.
Alternatively, expressing the effective phase $\delta_{h f \bar{f}}$ in terms of parameters of the two Higgs doublet theory using Eq. \eqref{eq:explicitphase}, 
limits may be presented directly in terms of the combination $|\lambda_5| \sin \textrm{Arg}\big( m_{12}^4 \lambda_5^*\big)\frac{v^2}{m_H^2}$.
The results are shown in  figure \ref{fig:plots:two}. 

We also present in  figure  \ref{fig:plots:three} the eEDM and the corresponding ACME II limits,
as a function of $\tan\beta$ and the imaginary part of the coupling modifier to the up-type quark Yukawas $\textrm{Im} \, \kappa_u$,
defined in Eq. \eqref{eq:effyukdef}.
The complex modifier of the up-type quark Yukawas is constrained to be at the per mille level or below in all types of 2HDM for all values of $\tan\beta$, 
except in a small region around $\tan\beta \simeq 1.3$ for types II and III, 
due to the cancellation in the contributions to the EDM mentioned above. 
Since the imaginary parts of the Higgs Yukawas share a universal phase and are related by powers of $\tan\beta$ for the types I-IV 2HDM,
the T-violating pieces of up, down and lepton Yukawas are correlated.
For this reason,
limits on $\textrm{Im} \, \kappa_u$ are due to both the top quark contribution to the eEDM, 
and to the contributions from the rest of the fermions.
As an example,
for the types II and III 2HDM, 
the bounds on $\textrm{Im} \, \kappa_u$ are strong at large $\tan\beta$ since for fixed $\textrm{Im} \, \kappa_u$, 
the lepton coupling modifier $\textrm{Im} \, \kappa_\ell$ is quadratically enhanced by $\tan\beta$, 
and leads to a large EDM. 
In addition, 
since in the 2HDM with Glashow-Weinberg conditions the coupling modifiers are family-universal,
these bounds apply to the imaginary part of the up, charm and top quark Yukawas.

In  figure \ref{fig:plots:four} we present the eEDM and the corresponding ACME II limits, 
as a function of $\tan\beta$ and the imaginary part of the complex coupling modifier to the lepton Yukawas $\textrm{Im} \, \kappa_\ell$.
Note that $\textrm{Im} \, \kappa_\ell$ is constrained to be below at the per mille level or below for all types of 2HDM and all values of $\tan\beta$, 
except in the region around $\tan\beta \simeq 1.3$ for the types II-III due to the cancellation discussed above. 

Finally and for completeness, 
in  figure \ref{fig:plots:five} we present the results for the eEDM as a function of $\tan\beta$ and $\textrm{Im} \, \kappa_d$,
the imaginary part of the complex coupling modifier of the down-type quark Yukawas.
We see that time-reversal violation in the down-type quark Yukawas is generically strongly constrained, 
except in the type IV 2HDM, 
where a large T-violating down-type Yukawa is still allowed for large values of $\tan\beta$.


\section{Conclusions}
\label{sec:five}

In this work, 
we studied the eEDM in the T-violating types I-IV two Higgs doublet theories near the decoupling limit,
using an effective theory approach.
We found that the leading contributions to the EDM come exclusively from Barr-Zee diagrams with effective dimension-six T-violating Higgs boson Yukawas generated by integrating out the heavy Higgs bosons.
In particular, 
contributions from heavy Higgs-mediated T-violating four-fermion interactions are GIM suppressed,
and arise only at effective dimension-eight.
This leads to a simplification of the analysis of the eEDM for all the types I-IV 2HDM near the decoupling limit,
allowing to express the EDM entirely in terms of two quantities:
$\tan\beta$ and a single effective dimension-six T-violating phase $\delta_{hf \bar{f}}$.
The dimension-six phase is defined by a unique combination of marginal and relevant Higgs potential couplings.
No additional individual reference to the heavy-Higgs masses or Higgs potential couplings is needed to obtain the eEDM at dimension six.

We presented limits on the effective dimension-six phase for the four types of 2HDM using the latest ACME II results,
and also presented bounds for the T-violating parts of the Higgs boson Yukawas.
We found that T violation in Higgs Yukawas for all the types I-IV 2HDM and for all fermions is constrained to be at the per mille level times the corresponding Standard Model Yukawa or below for all values of $\tan\beta$,
except in two cases. 
First, 
for a small region around $\tan\beta \simeq 1.3$ for the types II-III 2HDM, 
destructive interference between contributions to the eEDM allows for large T-violating Yukawas \cite{Bian:2014zka} to all fermions.
The second exception is in the type IV 2HDM, 
where large T-violating Yukawas to down-type quarks are still allowed for large values of $\tan\beta$.

Finally, 
in appendix \ref{app:four} we presented limits on T-violating Higgs boson Yukawa couplings to the electron, tau lepton, top and bottom quarks for generic theories beyond the types I-IV 2HDM.
We found that T violation in the top and electron Higgs Yukawas is constrained to be at or below the per mille level times the corresponding Standard Model Yukawa,
while T violation in the bottom and tau Higgs Yukawas is constrained to be below $\simeq 0.3$ times the corresponding Standard Model Yukawa.
 
In this work we did not discuss proposed future measurements of T violation in the Higgs sector at colliders \cite{Aad:2016nal,Chen:2017esh,Zanzi:2017msx,Harnik:2013aja,Berge:2013jra,Sun:2013yra,Anderson:2013afp,Chen:2013waa,Chen:2015gaa,Askew:2015mda,Belyaev:2015xwa,Buckley:2015vsa,Berge:2015nua,Bian:2017jpt,Chen:2017bff,Hagiwara:2016rdv,Li:2015kxc,Brehmer:2017lrt,Barberio:2017ngd,Han:2016bvf,Bernlochner:2018opw}.
These measurements could be complementary probes of T violation in Higgs-Yukawa interactions, 
especially in regions where destructive interference in the contribution to the eEDM from T violation in the different Higgs Yukawa interactions arises.


\bigskip
\bigskip

{ \Large \bf Acknowledgments}

\smallskip \smallskip

\noindent
This work was supported by the US Department of Energy 
under grant DE-SC0010008 and 
in part by the National Science Foundation under Grant No. NSF PHY-1748958.
The work of D.E.U. is supported by PHY-1620628.
The authors thank the Kavli Institute for Theoretical Physics, the Galileo Galilei Institute for Theoretical Physics and the INFN for the hospitality and partial support during the completion of this work. 


\appendix 
\section{Two-loop Dipole Functions} 

\label{app:one}
\begin{table}[ht]
\begin{center}
$
{\def\arraystretch{1.3}\tabcolsep=-10pt
\begin{array}{cc|cccccc}
 &  z 				& f(z)/\sqrt{z}    &  g(z)/\sqrt{z}   &   f(z)/z 		& g(z)/z    &  h(z)/z   & D(z)/z  \\ \hline
&  1 				&  0.83 		& 	1.17 		& 			& 		& -0.78	&	 -1.10			\\
m_t^2/m_h^2  & 1.90		&  0.73 		&      1.04		&			&		&		& 					\\
 m_b^2/m_h^2  & 1.11 \times 10^{-3} 	&  0.67 		&      0.83 		& 			& 		& 		& 				\\
 m_\tau^2/m_h^2 & 2.07 \times 10^{-4}	& 0.45  		&     0.54  		& 			& 		& 		&  				\\
 m_W^2/m_h^2 &  0.414 	& 	    		&			& 	1.47		& 	2.06	& -1.25	&  	-2.49			  \\
\end{array}}
$
\caption{ Numerical values for kinematic factors of the 
two-loop dipole functions 
defined in 
appendix \ref{app:one} evaluated at physical values
of the mass parameters. 
The kinematic factors are
equal to the two-loop functions 
with either a factor of the fermion mass or 
gauge boson mass squared factored out. 
The pole masses used in the kinematic factors are 
$m_t \simeq 172.5$ GeV, $m_b \simeq 4.2$ GeV, 
$m_\tau \simeq 1.8$ GeV, $m_W \simeq 80.4$ GeV, 
$m_h \simeq 125$ GeV. 
The $\overline{\rm MS}$ fermion masses used for the single overall 
factor proportional to  
Yukawa couplings
in the two-loop functions are 
$m_{t}(m_t) \simeq 163$ GeV, 
$m_{b}(m_h) \simeq 2.8$ GeV, 
$m_{\tau}(m_h) \simeq 1.7$ GeV. 
}
\label{tab:functions}
\end{center}
\end{table}

The two-loop dipole functions
$f(z), g(z)$ and $h(z)$ that appear in the expression for the 
electron electric dipole 
are \cite{Barr:1990vd,Leigh:1990kf}
\begin{eqnarray}
f(z)
&\! \! \! = \! \! \! &
\frac{z}{2} \, 
\int_{0}^1 dx\,
\frac{1-2x(1-x)}
{x(1-x)-z}
\ln
\Big[ \, 
\frac{x(1-x)}{z} \, 
\Big] \nonumber 
\\
g(z)
& \! \! \! = \! \! \! &
\frac{z}{2} \, 
\int_{0}^1 dx\,
\frac{1}
{x(1-x)-z}
\ln
\Big[ \, 
\frac{x(1-x)}{z}  \, 
\Big] \nonumber 
\\
h(z)
& \! \! \! = \! \! \! &
\frac{z}{2} \, 
\int_{0}^1 dx\,
\frac{1}
{x(1-x)-z}
\bigg(
\frac{z}
{x(1-x)-z}
\ln
\Big[ \, 
\frac{x(1-x)}{z} \,  
\Big]
-1
\bigg)
\end{eqnarray}
The two-loop function $D(z)$ is \cite{Leigh:1990kf}
\begin{equation}
D(z)=\sum_{i=1}^5 D_{i}(z)
\end{equation}
where  
\begin{eqnarray}
D_1(z)
& \! \! \! = \! \! \! &
-\frac{1}{2}
\int_{0}^1 dx 
\int_{0}^{1-x} dy \,
\frac{x}{B(x,y,z)}
\bigg(
\frac{2C(x,y,z)}{B(x,y,z)}
\big[ 3A(x,y,z)-2xy
\big]
-\bigg[
\frac{3a(x)-2xy}{a(x)}
\bigg]
\bigg)
\nonumber
\\
\nonumber
\\
\nonumber
\\
D_2(z)
&\! \! \! = \! \! \! &
\int_{0}^1   dx 
\int_{0}^{1-x} dy \,
x
\bigg( 
C'(x,y,z)
\bigg[
 \frac{3A(x,y,z)-2xy}{B^2(x,y,z)}
 +
 \frac{1+3x(1-2y)/(2a(x))}{B(x,y,z)}
 +\frac{3}{2a(x)}
 \bigg]
 \nonumber 
\\
 \nonumber 
\\
&&
\quad \quad\quad \quad \quad\quad \quad 
 +
 \,\,
 \frac{3A(x,y,z)-2xy}{2a(x)B(x,y,z)}
\, \bigg)
\nonumber \\
\nonumber
\\
\nonumber
\\
D_3(z)
& \! \! \! = \! \! \! &
\int_{0}^1 dx 
\int_{0}^{1-x} dy \,
\frac{x^2y}{a(x)(1-y-b)}
\bigg(
\frac{b(x,z)}{1-y-b(x,z)}
\ln\bigg[\frac{1-y}{b(x,z)}\bigg]
-1
\bigg)
\nonumber \\
\nonumber
\\
\nonumber
\\
D_4(z)
& \! \! \! = \! \! \! &
-\frac{1}{8}
\int_{0}^1 dx 
\int_{0}^{1-x} dy \,
\bigg(
\frac{1}{zB(x,y,z)}
\bigg[
1-\frac{2a(x) C(x,y,z)}{B(x,y,z)}
\bigg]
\nonumber \\
 \nonumber 
\\
& & \quad \quad \quad \quad \quad \quad
+
\,\,
\frac{x}{B(x,y,z)}
\bigg[
1-\frac{2A(x,y,z) C(x,y,z) }{B(x,y,z)}
\bigg]
\bigg)
\nonumber \\
\nonumber
 \\
 \nonumber
\\
D_5(z)
& \! \! \! = \! \! \! &
\frac{1}{8}
\int_{0}^1 dx 
\int_{0}^{1-x} dy \,
\frac{x}{a(x)}
\bigg(
\frac{C'(x,y,z)}{B^2(x,y,z)}
\bigg[x(2x-1) a(x)
+x(3x-1) B(x,y,z)
\nonumber \\ 
\nonumber \\ 
&&
\quad \quad \quad \quad \quad \quad \quad \quad
-
\,\,
2 B^2(x,y,z)
\bigg]
-2
\bigg[
1-
\frac{x(2x-1)}{4B(x,y,z)}
\bigg]  \, 
\bigg)
\end{eqnarray}
with 
\begin{eqnarray}
a(x)
& \! \! \! = \! \! \! &
x(1-x)
\nonumber \\
b(x,z) 
& \! \! \! = \! \! \! &
 a(x)/z
 \nonumber \\
A(x,y,z)
& \! \! \! = \! \! \! &
x+y/z 
\nonumber \\
B(x,y,z)
& \! \! \! = \! \! \! &
A(x,y,z)-a(x)
\nonumber \\
B'(x,y,z)
& \! \! \! = \! \! \! &
A(x,y,z)-a(y)
\nonumber \\ 
C(x,y,z)
& \! \! \! = \! \! \! &
\frac{A(x,y,z)}{B(x,y,z)}
\,
\ln \Big[ \, \frac{A(x,y,z)}{a(x)} \,  \Big]-1
\nonumber \\ 
C'(x,y,z)
& \! \! \! = \! \! \! &
\frac{a(x)}{B(x,y,z)}
\,
\ln \Big[ \, \frac{A(x,y,z)}{a(x)} \, \Big]-1
\end{eqnarray}


\section{Relation Between 2HDM Effective Theory and Unitary Mixing Languages} 
\label{app:two}
The $W$-boson two-loop eEDM diagrams in \cite{Leigh:1990kf} are expressed in the unitary mixing description proposed in \cite{Weinberg:1990me}.
In this appendix we provide the connection between the unitary mixing and effective field theory notations.
In the unitary mixing description, 
T violation in the 2HDM arises due to mixing of neutral Higgs bosons.
Reference \cite{Weinberg:1990me} defines neutral Higgs boson mixing function,
\begin{eqnarray}
A_0(q^2)&\equiv&\frac{2 e^{i(\xi_1-\xi_2)}}{v_1 v_2}\left<0|\Phi_2^0 \Phi_1^{0*} |0\right>
\quad ,
\label{eq:weinberg1}
\end{eqnarray}
where $v_1$ and $v_2$ are the condensates defined in Eq. \eqref{eq:defvev} and \eqref{eq:deftanb}.
Note that the phases $\xi_{1,2}$ are not gauge invariant. 
We work in the gauge
\begin{equation}
\xi_1=-\xi_2=-\frac{1}{2}\xi \quad \quad \quad {\textrm{(Gauge choice)}}
\end{equation}
Reference \cite{Weinberg:1990me} then defines the wave-function factors $Z_0$, $Z_0'$ and $Z_0''$ from the equality
\begin{eqnarray}
A_{0}(q^2)
&=&
\frac{1}{v^2}\frac{Z_{0}}{q^2+m_{\varphi _1}^2}
+
\frac{1}{v^2}\frac{Z_{02}}{q^2+m_{\varphi _2}^2}
+
\frac{1}{v^2}\frac{Z_{03}}{q^2+m_{\varphi _3}^2}
\quad ,
\label{eq:defZ}
\end{eqnarray}
where $\phi_{1,2,3}$ are the neutral Higgs mass eigenstates, 
ordered by increasing value of mass.
The $W$-boson two-loop eEDM diagrams in \cite{Leigh:1990kf} are expressed as functions of $\textrm{Im} \, Z_0$,
the wave-function factor of the lightest neutral Higgs.
We dedicate the rest of this appendix to find the relation between the wave-function factor $\textrm{Im} \, Z_0$ and our effective dimension-six T-violating phase $\delta_{h f \bar{f}}$ defined in Eq. \eqref{eq:dim6phase}.
The strategy will be to first find ${\rm Im } \, Z_0$ as a function of Higgs potential parameters, 
to then compare directly with expression \eqref{eq:dim6phase}.

The function ${\rm Im } \, Z_0$ is most easily obtained as a function of Higgs potential parameters using the Higgs basis.
We define Higgs basis doublets 
\begin{eqnarray}
e^{-i \xi /2} 
H_1 &=& \cos  \beta ~ \Phi_1+\sin \beta ~e^{-i\xi} ~\Phi_2 
\quad ,
  \nonumber \\
H_2 &=& -\sin \beta ~e^{i\xi} ~ \Phi_1+\cos \beta ~\Phi_2
\quad ,
\label{eq:rotationtoHiggsbasis}
\end{eqnarray}
such that in the new basis, only one of the doublets is responsible for EWSB
\begin{equation}
{v^2 \over 2} = \langle H_1^\dagger H_1 \rangle  
\quad , \quad
0 = \langle H_2^\dagger H_2 \rangle 
\quad .
\label{eq:HiggscondensatesHiggsbasis}
\end{equation}
The Higgs basis potential is 
\begin{eqnarray}
 V(H_1,H_2) &=&{\tilde{m}_1}^2  H_1^\dagger  H_1+{\tilde{m}_2}^2  H_2^\dagger  H_2+\Big({\tilde{m}_{12}}^2  H_1^\dagger  H_2 +\textrm{h.c.}\Big) \nonumber  \nonumber \\
&+&\frac{1}{2} {\tilde{\lambda}}_1 ( H_1^\dagger  H_1)^2+ \frac{1}{2} {\tilde{\lambda}}_2( H_2^\dagger  H_2)^2+ {\tilde{\lambda}}_3( H_2^\dagger  H_2)( H_1^\dagger  H_1)+ {\tilde{\lambda}}_4 ( H_2^\dagger  H_1)( H_1^\dagger  H_2) \nonumber  \nonumber \\
&+&\bigg[ ~ \frac{1}{2}{\tilde{\lambda}}_5( H_1^\dagger  H_2)^2+{\tilde{\lambda}}_6  H_1^\dagger  H_1  H_1^\dagger  H_2 +{\tilde{\lambda}}_7( H_2^\dagger  H_2)( H_1^\dagger  H_2)+\textrm{h.c.}~\bigg]
\quad .
\label{eq:2HDMLagrangian}
\end{eqnarray}
We also define Higgs basis fields
\begin{eqnarray} 
H_1^0 \! \! \! &=& \! \! \! {1 \over \sqrt{2}} \Big( v + h_1 + i G^0 \Big)
\quad , 
\nonumber \\
H_2^0 \! \! \! &=&  \! \! \! {1 \over \sqrt{2}} 
     ~ e^{ i~ \! \! {\rm Arg}({\tilde{\lambda}}_5^*) /2}    
          \Big( h_2 + i h_3 \Big)
          \quad ,
  \nonumber
\\
H_1^+ \! \! \! &=&  \! \! \! G^+ 
\quad ,
\nonumber \\
H_2^+ \! \! \! &=&  \! \! \! 
      e^{ i~ \! \! {\rm Arg}({\tilde{\lambda}}_5^*) /2}    
        ~ H^+ 
        \quad .
    \label{eq:Higgsbasis}
 \end{eqnarray}
 The charged Higgs does not mediate T violation in the unitary mixing description \cite{Weinberg:1990me},
 so we concentrate on the neutral Higgs bosons.
The massive neutral components of the doublets in the original basis can be expressed in term of neutral Higgs basis fields Eq. \eqref{eq:Higgsbasis},
\begin{eqnarray}
 \Phi_a^0 
 &=& 
 N_{a j} \, h_j 
 \quad \quad \quad  a=1,2 \quad , \quad i=1..3
 \quad ,
\end{eqnarray}
where $N_{aj}$ is a non-unitary matrix fixed by Eq. \eqref{eq:rotationtoHiggsbasis} and \eqref{eq:Higgsbasis},
\begin{equation}
\begin{array}{ccccccc}
N_{11}
&=&
\frac{1}{\sqrt{2}} 
\, e^{-i\xi/2} \cos\beta 
&&
N_{21}
&=&
\frac{1}{\sqrt{2}} 
\, e^{i\xi/2} \sin \beta 
\\
N_{12}
&=&
-\frac{1}{\sqrt{2}}
e^{i~ \! \! {\rm Arg}({\tilde{\lambda}}_5^*) /2}
e^{-i\xi} 
\sin\beta
&&
N_{22}
&=&
\frac{1}{\sqrt{2}}
e^{i~ \! \! {\rm Arg}({\tilde{\lambda}}_5^*) /2}
e^{-i\xi} 
\cos\beta \\
N_{13}
&=&
-\frac{1}{\sqrt{2}}
i \, e^{i~ \! \! {\rm Arg}({\tilde{\lambda}}_5^*) /2}
e^{-i\xi} 
\sin\beta
&&
N_{23}
&=&
\frac{1}{\sqrt{2}}
i \, e^{i~ \! \! {\rm Arg}({\tilde{\lambda}}_5^*) /2}
e^{-i\xi} 
\cos\beta
\end{array}
\quad \quad .
\label{eq:Nmatrix}
\end{equation}
The neutral massive eigenstates $\varphi_{1,2,3}$ are combinations of the Higgs basis neutral states $h_{1,2,3}$.
We identify the lightest mass eigenstate $h\equiv\varphi_1$ with the 125 GeV Higgs. 
The Higgs basis neutral states in terms of the massive eigenstates are
\begin{equation}
h_i = \widehat{V}_{i}^a\, \varphi_a
\quad ,
\end{equation}
where the matrix $\widehat{V}$ is given in \cite{Egana-Ugrinovic:2015vgy}. 
In particular, 
the projections of Higgs basis states into the Higgs $\varphi_1$ are given by the complex alignment parameter $\Xi$ \cite{Egana-Ugrinovic:2015vgy}
\begin{eqnarray}
\widehat{V}^1_1=\sqrt{1-|{\Xi }| ^2}  ~~~ ~~~ \widehat{V}^1_2=\textrm{Re}~\!\Xi  ~~~ ~~~ \widehat{V}^1_3=\textrm{Im}~\!\Xi  ~~~,  ~~~ \Xi  \in \mathbb{C}^1
\quad .
\label{eq:Vh}
\end{eqnarray}
The complex alignment parameter at lowest order in an expansion of the electroweak scale over the heavy Higgs mass scale is given by \cite{Egana-Ugrinovic:2015vgy}
\begin{equation}
\Xi ~\!  e^{ i ~ \! \! {\rm Arg}({\tilde{\lambda}}_5^*)/2}=-{{\tilde{\lambda}}_6^*} ~\! \frac{v^2}{m_H^2}+{\cal O}\bigg({\lambda^3 v^4 \over m_H^4}\bigg)
\quad .
\label{eq:convenientxi}
\end{equation}
We now express the correlation functions Eq. \eqref{eq:weinberg1} in terms of projections into neutral massive eigenstates, 
\begin{eqnarray}
A_0(q^2)
&=&
\frac{2 e^{-i\xi}}{v^2 \sin\beta\cos\beta}
~ \frac{1}{q^2+m_{\varphi_a}^2} ~ 
\Big[ \widehat{V}^a_i 
N_{2i}
\Big]
\Big[
\widehat{V}^a_j
N_{1j}
\Big]^*
\quad .
\label{eq:projections}
\end{eqnarray}
where we sum over repeated indices.  
Here we consider only T violation in the propagation of the lightest mass eigenstate, 
which is all the source of T violation in the low energy theory at effective dimension-six as shown in section \ref{sec:three}. 
Comparing Eq. \eqref{eq:projections} and \eqref{eq:defZ} we obtain
\begin{eqnarray}
Z_{0}&=& 
\frac{2 e^{-i\xi}}{\sin\beta\cos\beta}
\,
\Big[ \widehat{V}^1_i 
N_{2i}
\Big]
\Big[
\widehat{V}^1_j
N_{1j}
\Big]^*
\quad .
\label{eq:Z0}
\end{eqnarray}
Using Eq. \eqref{eq:Nmatrix}, \eqref{eq:Vh} and \eqref{eq:convenientxi} in \eqref{eq:Z0} we obtain 
\begin{eqnarray} 
{\rm Im } \, Z_0
&=&
{\rm Im}
\bigg[
e^{-\frac{i}{2}\xi} e^{\frac{i}{2} {\rm Arg}(\tilde{\lambda}_5^*) }~\!\Xi   \sqrt{1-|{\Xi}|^2} \cot \beta
-
e^{\frac{i}{2}\xi} e^{-\frac{i}{2} {\rm Arg}(\tilde{\lambda}_5^*) }~\!\Xi   \sqrt{1-|{\Xi}|^2} \tan \beta
\bigg]
\nonumber 
\\
&=&
-
 \frac{v^2}{m_H^2}~\!
{\rm Im}
\bigg[
 e^{-i \xi/2} {\tilde{\lambda}}_6^* \cot \beta
 -
  e^{i \xi/2} {\tilde{\lambda}}_6 \tan \beta
 \bigg]
 +{\cal O}\bigg({\lambda^3 v^4 \over m_H^4}\bigg)
 \quad .
 \label{eq:ImZ1temp}
 \end{eqnarray}
The Higgs basis coupling $\tilde{\lambda}_6$ in the above expression may be rewritten in terms of Higgs potential couplings of the original Glashow-Weinberg Higgs doublet basis,
using the basis transformation relation Eq. \eqref{eq:rotationtoHiggsbasis}.
The result is \cite{Egana-Ugrinovic:2015vgy}
\begin{equation}
 {\tilde{\lambda}}_6 e^{i\xi/2}
 =
 -\frac{1}{2}\sin 2\beta  \left(\lambda_1 \cos^2 \beta-\lambda_2 \sin^2{\beta}-\lambda_{345}\cos 2\beta -i\textrm{Im}(\lambda_5 e^{2i\xi})\right)
 \label{eq:lambda6}
 \quad .
\end{equation}
Using Eq. \eqref{eq:lambda6} in \eqref{eq:ImZ1temp} we obtain
\begin{eqnarray} 
{\rm Im } \, Z_0
&=&
\frac{1}{2} (\cot \beta+\tan\beta) \sin{2\beta}
 \frac{v^2}{m_H^2}~\!
\textrm{Im}(\lambda_5 e^{2i\xi})
 +{\cal O}\bigg({\lambda^3 v^4 \over m_H^4}\bigg)
 \nonumber \\
&=&
- (\cot \beta+\tan\beta)~\sin~\!\delta_{h f \bar{f}}+{\cal O}\bigg({\lambda^3 v^4 \over m_H^4}\bigg)
\quad .
\label{eq:finalZ0}
\end{eqnarray}
where in the last line we made use of the definition of the effective dimension-six phase $\delta_{h f \bar{f}}$ in Eq. \eqref{eq:dim6phase}.

Alternatively, 
we may express ${\rm Im } \, Z_0$ in terms of the complex coupling modifiers of the lepton Yukawas, 
which are responsible for T violation in the W boson mediated two-loop diagrams.
Since in \cite{Leigh:1990kf} the $W$-boson mediated EDM diagrams are calculated for the type II 2HDM,
we must use the relation $\textrm{Im} \, \kappa_{\ell}=\tan\beta \sin~\!\delta_{h f \bar{f}} $ from table \ref{tab:Yukawas}.
We obtain
\begin{eqnarray}
\textrm{Im} \, Z_0 
&=&
-\frac{ \textrm{Im} \,{\kappa}_\ell}{\sin^2\beta}
+{\cal O}\Big({\lambda^3 v^4 \over m_H^4}\Big)
\quad .
\label{eq:appBfinalresult}
\end{eqnarray}
The $W$-boson two-loop eEDM diagrams for the rest of the two Higgs doublet theories (types I,III and IV), 
may then be simply obtained by replacing $ \textrm{Im} \,{\kappa}_\ell$ by the corresponding expression from table \ref{tab:Yukawas}.

\section{Effective dimension-six phase in terms of Higgs potential phases}
\label{app:three}
In this appendix we derive an expression for the effective dimension-six phase defined in Eq. \eqref{eq:dim6phase} purely in terms of Higgs potential parameters, 
by solving for the condensate phase $\xi$ using the Higgs potential minimization conditions.
The Higgs potential Eq. \eqref{eq:2HDMpotentialgeneric} evaluated in the neutral vacuum defined in Eq. \eqref{eq:defvev} is given by
\begin{eqnarray}
V(v_1 , v_2, \xi) \! \! \! &=& \! \! \! {1 \over 2} m_1^2 v_1^2
+ {1 \over 2} m_2^2 v_2^2
+ {1 \over 8} \lambda_1 v_1^4 
+ {1 \over 8} \lambda_2 v_2^4 
+ {1 \over 4} \big( \lambda_3+ \lambda_4) v_1^2 v_2^2 
 \nonumber \\
&&+~  \! {\rm Re} \Big( 
 -  v_1 v_2 m_{12}^2 e^{i \xi} + 
 {1 \over 4} v_1^2 v_2^2 \lambda_5 e^{2i \xi} 
 \Big)
 \quad .
  \label{eq:condensatepotential}
\end{eqnarray}
The minimization condition of the condensate potential Eq. \eqref{eq:condensatepotential} for the effective phase $\xi$ is
\begin{eqnarray}
\frac{\partial V}{\partial \, \xi}
&=&
0\,\,=\,\,
\frac{1}{2 }v^2 \sin 2\beta\,  \, \big[\, \textrm{Im} (m_{12}^2 e^{i \xi})
-\frac{1}{4} v^2 \sin 2\beta \, \textrm{Im} ( \lambda_5 e^{2i\xi})
\big]
\quad ,
\label{eq:EWSB1}
\end{eqnarray}
or equivalently, for $ \sin 2\beta \neq 0$
\begin{eqnarray}
\textrm{Im}\, (m_{12}^{2} e^{i\xi}\big)
&=&
\frac{1}{4} v^2\, \textrm{Im} ( \lambda_5 e^{2i\xi})  \sin 2\beta 
\quad .
\label{eq:equivalence2}
\end{eqnarray}
Dividing Eq. \eqref{eq:equivalence2} by $|{m_{12}}|^2$, we obtain
\begin{equation}
\textrm{Im} \big(e^{i \textrm{Arg}\,m_{12}^2} e^{i \xi}\big)
=
\frac{1}{4} \frac{v^2}{|m_{12}|^2} \, \textrm{Im} ( \lambda_5 e^{2i\xi})\, \sin 2 \beta
\quad .
\label{eq:eq:EWSB1p}
\end{equation}
In the unitary mixing language, the mass term $|m_{12}^2|$ is related to the heavy Higgs masses through \cite{Egana-Ugrinovic:2015vgy}
\begin{equation}
|m_{12}^2| 
=
 m_H^2
  \, \bigg[\frac{1}{2}\sin 2\beta+{\cal O}\bigg(\frac{\lambda  v^2}{m_H^2}\bigg)\bigg]
  \quad ,
\label{eq:m12}
\end{equation}
so to lowest order in $v^2/m_H^2$ the EWSB condition Eq. \eqref{eq:eq:EWSB1p} may be rewritten as
\begin{equation}
\textrm{Im} \big(e^{i \textrm{Arg}\,m_{12}^2} e^{i \xi}\big)
=
\frac{1}{2} \frac{v^2}{m_H^2} \,\textrm{Im}\big( \lambda_5 e^{2i\xi} \big)
\bigg[
1+
{\cal O}\bigg(
\,
\frac{\lambda v^2}{m_H^2}
\,
\bigg)
\bigg]
\quad ,
\label{eq:EWSB2p}
\end{equation}
or equivalently
\begin{equation}
\sin \big(\textrm{Arg}\,m_{12}^2+{i \xi}\big)
=
\frac{1}{2} |\lambda_5| \frac{v^2}{m_H^2} \sin \big(\textrm{Arg}\,\lambda_5+{2i \xi}\big)
\bigg[
1+
{\cal O}\bigg(
\,
\frac{\lambda v^2}{m_H^2}
\,
\bigg)
\bigg]
\quad ,
\label{eq:EWSB3p}
\end{equation}
so we find
\begin{equation}
\xi
=
-\textrm{Arg}\,m_{12}^2
+
{\cal O}\bigg(
\,
 \frac{\lambda v^2}{m_H^2}
\,
\bigg)
\quad .
\label{eq:EWSB2p}
\end{equation}
Using Eq. \eqref{eq:EWSB2p} in the effective dimension-six phase Eq. \eqref{eq:dim6phase} we obtain the  effective phase in terms of Higgs potential parameters, 
\begin{eqnarray}
\sin \delta_{h f \bar{f}}
&\equiv&
-
  \frac{1}{2}\, 
\sin (2\beta)
\,
\textrm{Im}(\lambda_5 e^{2i\xi})
\frac{v^2}{m_H^2} 
\nonumber \\
&=&
-
  \frac{1}{2}\, 
\sin (2\beta)
\,
\textrm{Im}(\lambda_5 e^{-2i \textrm{Arg}\,m_{12}^2})
\frac{v^2}{m_H^2} 
\bigg[
1+
{\cal O}\bigg(
\,
\frac{\lambda v^2}{m_H^2}
\,
\bigg)
\bigg]
\nonumber \\
&=&
  \frac{1}{2}|\lambda_5|\, 
\sin (2\beta)
\sin \textrm{Arg}\big( m_{12}^4 \lambda_5^*\big)
\frac{v^2}{m_H^2} 
\bigg[
1+
{\cal O}\bigg(
\,
\frac{\lambda v^2}{m_H^2}
\,
\bigg)
\bigg]
\quad .
\label{eq:dim6phaseapp}
\end{eqnarray}


\section{Bounds on generic theories with P- and T-violating Higgs Yukawa interactions}
\label{app:four}

As discussed in section \ref{sec:four},
keeping track of correlations between the Yukawa interactions to different fermions is important for calculating the eEDM in the types I-IV 2HDM. 
Since the correlations are specific to each type of 2HDM, 
the limits presented above are not easily applicable to other theories with T-violating Higgs-Yukawa interactions.
To cover other theories beyond the types I-IV 2HDM,
in this appendix we calculate the eEDM and set limits on generic theories with T violation in individual Higgs Yukawa interactions.
For concreteness, 
in this section we allow only for T violation in the Higgs interactions with the electron and with third-generation fermions.
We parametrize T violation in Higgs fermion interactions using complex coupling modifiers.
In the fermion mass eigenbasis,
the complex modifiers $\kappa_f$ are defined as
\begin{eqnarray}
 \lambda^{f}_{h  ij}
  \! \! \! &=& \! \! \!
\delta_{ij}~\!\frac{m_{i}^{f}}{v}
\, (\, 1+\kappa_{f} \, )
\quad ,
\label{eq:effyukdef2}
\end{eqnarray}
where in $f=e,\tau,t,b$ and the Yukawa couplings are defined with the conventions in Eq. \eqref{eq:yukawadef}.
The phase of the complex coupling modifier is a physical T-violating phase,
since it measures the relative phase between the Yukawa coupling and the corresponding fermion mass, 
which here is taken to be real. 
Differently from the case of the types I-IV 2HDM discussed in the body of this paper,
here the coupling modifiers to the different Standard Model fermions are independent parameters.
The complex modifiers may arise from a short-distance dimension-six cubic Higgs Yukawa operator as in \cite{Altmannshofer:2015qra,Zhang:1994fb,Bodeker:2004ws}.
Since we are interested in setting limits on the T-violating piece of the complex coupling modifiers, 
for simplicity we assume that $\textrm{Re} \, \kappa_{f}=0$. 
In the case of third-generation fermions, this assumption is also an approximation to the currently measured values of overall coupling modifiers \cite{ATLAS:2018doi,Aaboud:2018urx,ATLAS:2018nkp,ATLAS:2018lur,CMS:2018lkl,Sirunyan:2018hoz,CMS:2018nqp}.

The leading contributions of the above T-violating Higgs Yukawa interactions to the eEDM come from two types of Barr-Zee diagrams.
For T violation in the electron Yukawa, 
the leading contributions arise from two-loop diagrams with third-generation fermions in the loop, 
as in figure \ref{fig:BarrZee_D6_3rdgen} (left), 
or from diagrams with $W$-bosons (or ghosts), 
as in figure \ref{fig:BarrZee_D6}.
For T violation in the Higgs interactions with the $\tau$ lepton, top or bottom quarks, 
the leading contributions come from diagrams with the corresponding fermion in the loop, 
as illustrated in figure \ref{fig:BarrZee_D6_3rdgen} (right).
The eEDM from these diagrams in terms of T-violating coupling modifiers is
\begin{eqnarray}
\frac{d_e}{e} 
& \! \! \! \simeq \! \! \! &
 \frac{2 \sqrt{2} \, \alpha \, G_F m_e}{(4\pi)^3} \bigg[ \, 
 -
\Big(
 - 
 \frac{16}{3}
 f(m_t^2/m_h^2)
 - \frac{4}{3}
 f(m_b^2/m_h^2)
 -4 \, 
f(m_\tau^2/m_h^2)
 \nonumber \\ 
 &&
\hspace*{2cm}  +~~
2 
~
\bigg[ 
~\! 
5g(m_W^2/m_h^2)+3 f(m_W^2/m_h^2) 
+
\frac{3}{4} \, 
\Big[
\,
g(m_W^2/m_h^2)+h(m_W^2/m_h^2)
\,
\Big]
~\! 
\nonumber \\
&&
\hspace*{2cm}-~~
\frac{g(m_W^2/m_h^2)-f(m_W^2/m_h^2)}{2 \, m_W^2/m_h^2}
\, \bigg]
+
\frac{1}{2\sin^2 \theta_W }
\,
D(m_W^2/m_h^2)
\,
\Big)
~
 \textrm{Im}\, \kappa_e
 \nonumber \\
 &&
\hspace*{2cm} - ~~
 \Big( \, 
 \frac{16}{3}
 g(m_t^2/m_h^2)
 \,
 \textrm{Im}\, \kappa_t
 - \frac{4}{3}
 g(m_b^2/m_h^2)
 \,
 \textrm{Im}\, \kappa_b
 - 4
 g(m_\tau^2/m_h^2)
 \,
  \textrm{Im}\, \kappa_\tau
 \,
 \Big)
 ~  
 ~
 \bigg]
 \nonumber
 \quad .
 \\
 \label{edmformula2}
\end{eqnarray}
Comparing Eq. \eqref{edmformula2} with the ACME  II limit Eq. \eqref{eq:experimentalbound}, 
we set limits on the T-violating electron, tau, top and bottom Yukawa interactions.
We first set limits allowing for only one of the Yukawa coupling modifiers to be non-zero at a time. 
We obtain
\begin{eqnarray}
|\textrm{Im} \, \kappa_e| &<& 2.0 \times 10^{-3} 
\nonumber \\
|\textrm{Im} \, \kappa_\tau| &<& 3.0  \times 10^{-1}
\nonumber \\
|\textrm{Im} \, \kappa_t| &<& 1.3  \times 10^{-3} 
\nonumber \\
|\textrm{Im} \, \kappa_b| &<& 3.7  \times 10^{-1} \quad \quad .
\end{eqnarray}
T violation in the electron and top Yukawa is strongly constrained to be at or below the per mille level times the corresponding Standard Model Yukawa. 
Our limits are consistent with the ones presented in \cite{Altmannshofer:2015qra,Brod:2013cka} using the previous ACME limit \cite{Baron:2013eja}.
The limits on the tau and bottom T-violating Yukawa  modifiers are weaker,
since the tau and bottom contributions to the eEDM come from the diagrams in figure \ref{fig:BarrZee_D6_3rdgen} (left),
which are suppressed by small tau and bottom mass insertions.

\begin{figure}[h]
\begin{center}
\vspace*{.2in}
\includegraphics[width=16cm]{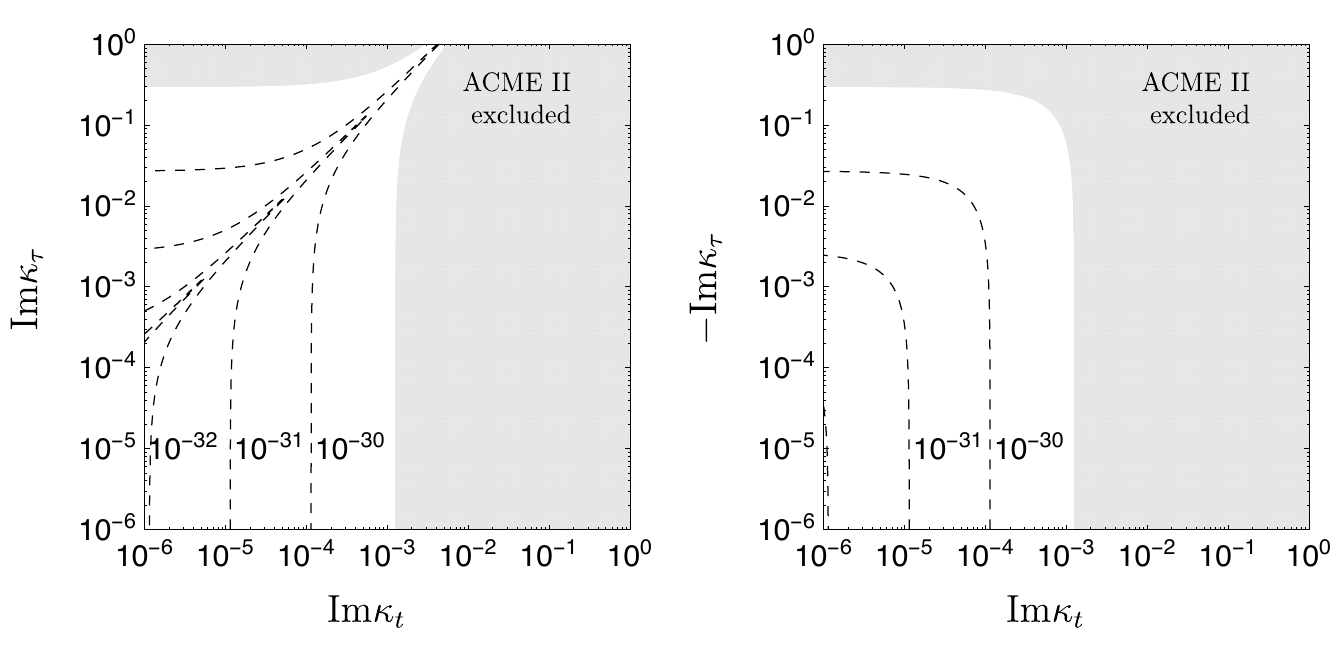}
\caption{
Contributions to the 
magnitude of the electron electric dipole moment in units of $e$ cm from a theory with P- and T-violating Higgs Yukawa couplings to the top quark and tau lepton.
$ \kappa_{t,\tau}$ are coupling modifiers defined in Eq. \eqref{eq:effyukdef2}, 
which are assumed to be purely imaginary. 
The rest of the Higgs couplings to fermions and gauge bosons are set to their Standard Model values. 
The shaded region is excluded at $90\% \, \textrm{CL}$ by the current bound on the electron electric dipole moment of 
$|d_e| < 1.1 \times 10^{-29}$ $e$ cm from the ACME II experiment \cite{Andreev:2018ayy}. 
}
\label{fig:plots:six}
\end{center}
\end{figure}

\begin{figure}[h]
\begin{center}
\vspace*{.2in}
\includegraphics[width=16cm]{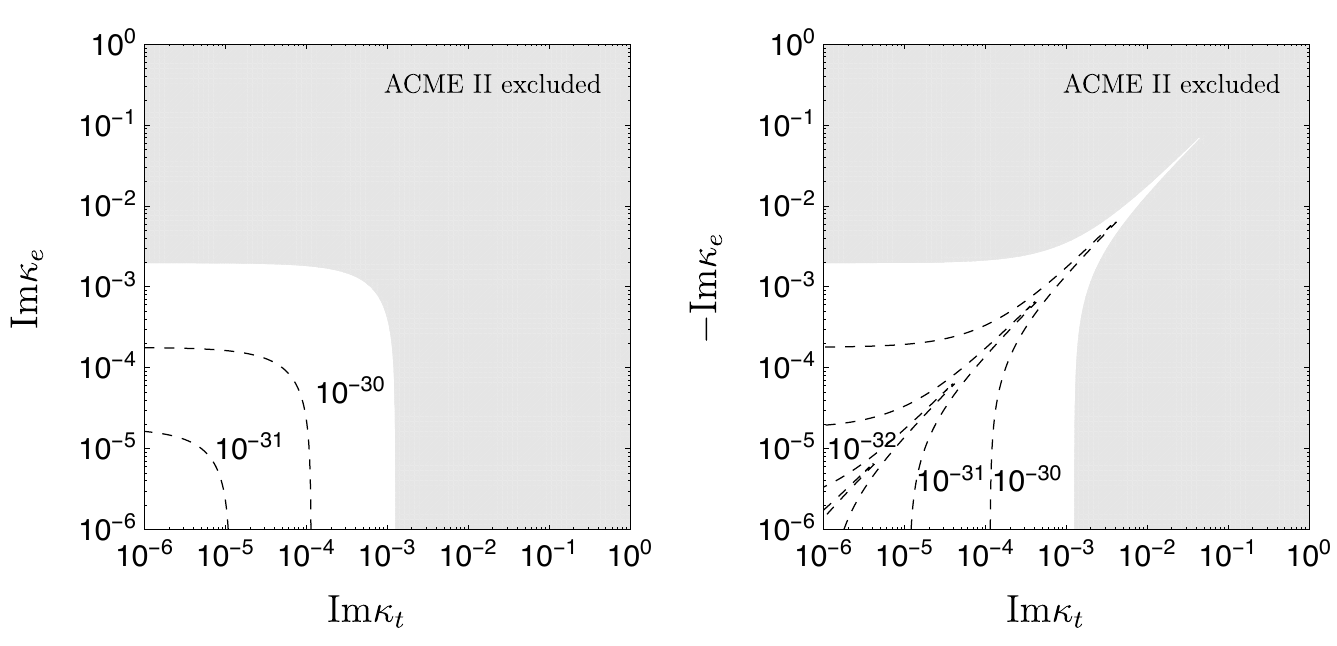}
\caption{Contributions to the 
magnitude of the electron electric dipole moment in units of $e$ cm from a theory with P- and T-violating Higgs Yukawa couplings to the top quark and electron.
$\kappa_{t,e}$ are coupling modifiers defined in Eq. \eqref{eq:effyukdef2}, 
which are assumed to be purely imaginary. 
The rest of the Higgs couplings to fermions and gauge bosons are set to their Standard Model values. 
The shaded region is excluded at $90\% \, \textrm{CL}$ by the current bound on the electron electric dipole moment of 
$|d_e| < 1.1 \times 10^{-29}$ $e$ cm from the ACME II experiment \cite{Andreev:2018ayy}. 
}
\label{fig:plots:seven}
\end{center}
\end{figure}

\begin{figure}[h]
\begin{center}
\vspace*{.2in}
\includegraphics[width=16cm]{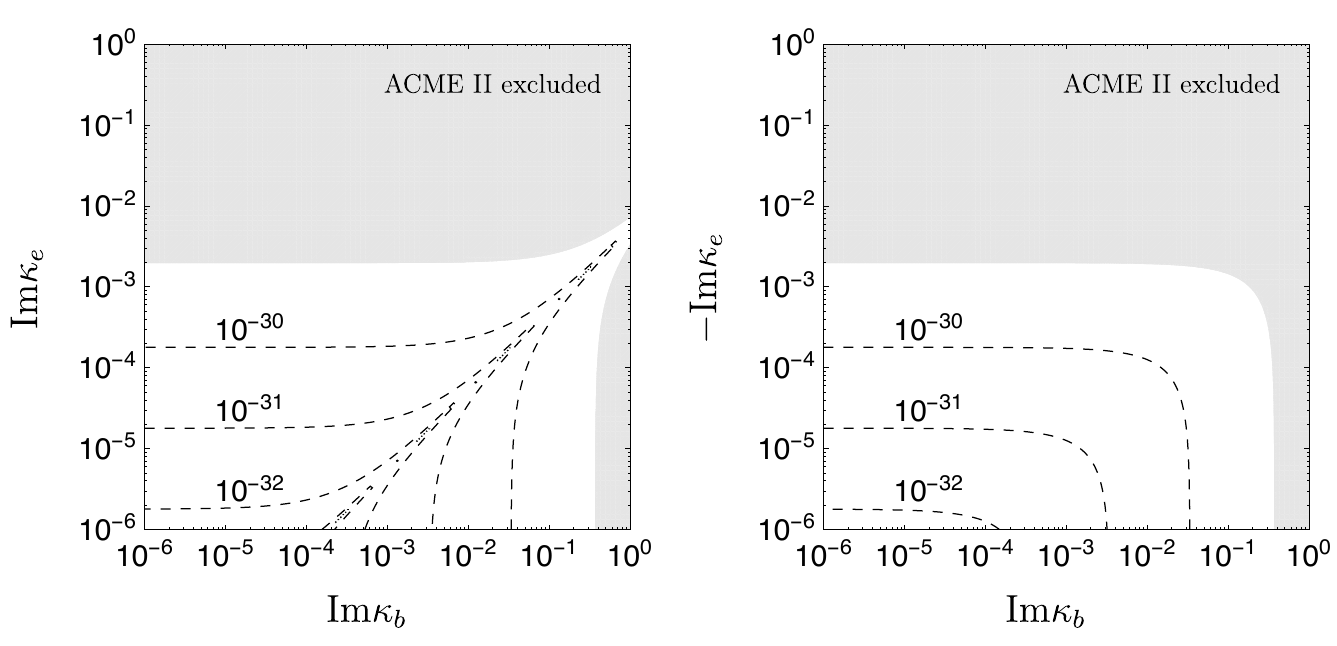}
\caption{Contributions to the 
magnitude of the electron electric dipole moment in units of $e$ cm from a theory with P- and T-violating Higgs Yukawa couplings to the bottom quark and electron.
$\kappa_{b,e}$ are coupling modifiers defined in Eq. \eqref{eq:effyukdef2}, 
which are assumed to be purely imaginary. 
The rest of the Higgs couplings to fermions and gauge bosons are set to their Standard Model values. 
The shaded region is excluded at $90\% \, \textrm{CL}$ by the current bound on the electron electric dipole moment of 
$|d_e| < 1.1 \times 10^{-29}$ $e$ cm from the ACME II experiment \cite{Andreev:2018ayy}. 
}
\label{fig:plots:eight}
\end{center}
\end{figure}

The limits change when two or more T-violating Higgs Yukawa interactions are non-zero,
since destructive interference between different contributions to the eEDM may arise.
Consider for instance
allowing for T violation in both the top and tau Yukawa interactions. 
Contours of the eEDM as a function of $\textrm{Im} \, \kappa_t$ and $\textrm{Im} \, \kappa_\tau$ and the corresponding ACME II limits are presented in figure \ref{fig:plots:six}.
In  figure  \ref{fig:plots:six} (left), 
we present the results considering the same sign for $\textrm{Im} \, \kappa_\tau$ and $\textrm{Im} \, \kappa_t$,
whereas in  figure \ref{fig:plots:six} (right) we consider the opposite sign.
In the former case, 
destructive interference is possible \footnote{We remind the reader that our Yukawa conventions Eq. \eqref{eq:yukawadef} introduce a sign difference in the imaginary part of the up-type quark and lepton Yukawas}.
An exact cancellation between the top and tau contributions to the eEDM happens around $\textrm{Im} \, \kappa_\tau \simeq 2.4  \times 10^2 \, \textrm{Im} \, \kappa_t$.
In this region, 
T violation in the top and tau Higgs Yukawa interactions cannot be constrained by limits on the eEDM.
On the other hand, 
if we allow for T violation in the top and electron Yukawa modifiers instead,
the results are presented in figure  \ref{fig:plots:seven} for the two possible choices of relative signs between the two Yukawa modifiers. 
In this case, 
due to a difference in sign in the relevant diagrams for electron and top T-violating Yukawas,
destructive interference arises when the sign of the imaginary part of these two Yukawas is the opposite. 
A complete cancellation between the two contributions happens around  $\textrm{Im} \, \kappa_e \simeq - 1.6 \, \textrm{Im} \, \kappa_t$.
Finally, 
in figure \ref{fig:plots:eight} we allow for T violation in the bottom and electron Yukawas. 
A complete cancellation of the contributions to the eEDM occurs for $\textrm{Im} \, \kappa_b \simeq 1.9 \times 10^2 \, \textrm{Im} \, \kappa_e$.

\begin{thebibliography}{100}

\bibitem{Aad:2012tfa} G.~Aad {\it et al.} [ATLAS Collaboration],
  ``Observation of a new particle in the search for the Standard Model Higgs boson with the ATLAS detector at the LHC,''
  Phys.\ Lett.\ B {\bf 716}, 1 (2012)
  [arXiv:1207.7214 [hep-ex]].

\bibitem{Chatrchyan:2012xdj} S.~Chatrchyan {\it et al.} [CMS Collaboration],
  ``Observation of a new boson at a mass of 125 GeV with the CMS experiment at the LHC,''
  Phys.\ Lett.\ B {\bf 716}, 30 (2012)
  [arXiv:1207.7235 [hep-ex]].

\bibitem{Aad:2015gba} G.~Aad {\it et al.} [ATLAS Collaboration],
  Eur.\ Phys.\ J.\ C {\bf 76}, no. 1, 6 (2016)
  [arXiv:1507.04548 [hep-ex]].

\bibitem{Khachatryan:2014jba} V.~Khachatryan {\it et al.} [CMS Collaboration],
  ``Precise determination of the mass of the Higgs boson and tests of compatibility of its couplings with the standard model predictions using proton collisions at 7 and 8 $\,\text {TeV}$,''
  Eur.\ Phys.\ J.\ C {\bf 75}, no. 5, 212 (2015)
  [arXiv:1412.8662 [hep-ex]].

\bibitem{combination} The ATLAS and CMS Collaborations,
  ``Measurements of the Higgs boson production and decay rates and constraints on its couplings from a combined ATLAS and CMS analysis of the LHC pp collision data at $\sqrt{s}$ = 7 and 8 TeV,''
  ATLAS-CONF-2015-044.
  
\bibitem{ATLAS:2018doi} 
  The ATLAS collaboration [ATLAS Collaboration],
  ATLAS-CONF-2018-031.
  
\bibitem{Aaboud:2018urx} 
  M.~Aaboud {\it et al.} [ATLAS Collaboration],
  Phys.\ Lett.\ B {\bf 784}, 173 (2018)
  doi:10.1016/j.physletb.2018.07.035
  [arXiv:1806.00425 [hep-ex]].
  
\bibitem{ATLAS:2018nkp} 
  The ATLAS collaboration [ATLAS Collaboration],
  ATLAS-CONF-2018-036.
  
\bibitem{ATLAS:2018lur} 
  The ATLAS collaboration [ATLAS Collaboration],
  ATLAS-CONF-2018-021.
  
\bibitem{CMS:2018lkl} 
  CMS Collaboration [CMS Collaboration],
  CMS-PAS-HIG-17-031.
  
\bibitem{Sirunyan:2018hoz} 
  A.~M.~Sirunyan {\it et al.} [CMS Collaboration],
  Phys.\ Rev.\ Lett.\  {\bf 120}, no. 23, 231801 (2018)
  doi:10.1103/PhysRevLett.120.231801, 10.1130/PhysRevLett.120.231801
  [arXiv:1804.02610 [hep-ex]].
  
\bibitem{CMS:2018nqp} 
  CMS Collaboration [CMS Collaboration],
  CMS-PAS-HIG-18-007.
  
\bibitem{Andreev:2018ayy} 
  V.~Andreev {\it et al.} [ACME Collaboration],
  Nature {\bf 562}, no. 7727, 355 (2018).
  doi:10.1038/s41586-018-0599-8
  


\bibitem{McLerran:1990zh} L.~D.~McLerran, M.~E.~Shaposhnikov, N.~Turok and M.~B.~Voloshin,
  Phys.\ Lett.\ B {\bf 256}, 451 (1991).
  doi:10.1016/0370-2693(91)91794-V

\bibitem{Turok:1990zg} N.~Turok and J.~Zadrozny,
  Nucl.\ Phys.\ B {\bf 358}, 471 (1991).
  doi:10.1016/0550-3213(91)90356-3

\bibitem{Cohen:1991iu} A.~G.~Cohen, D.~B.~Kaplan and A.~E.~Nelson,
  Phys.\ Lett.\ B {\bf 263}, 86 (1991).
  doi:10.1016/0370-2693(91)91711-4

\bibitem{Zhang:1994fb} X.~Zhang, S.~K.~Lee, K.~Whisnant and B.~L.~Young,
  Phys.\ Rev.\ D {\bf 50}, 7042 (1994)
  doi:10.1103/PhysRevD.50.7042
  [hep-ph/9407259].

\bibitem{Cline:1995dg} J.~M.~Cline, K.~Kainulainen and A.~P.~Vischer,
  Phys.\ Rev.\ D {\bf 54}, 2451 (1996)
  doi:10.1103/PhysRevD.54.2451
  [hep-ph/9506284].

\bibitem{Bodeker:2004ws} D.~Bodeker, L.~Fromme, S.~J.~Huber and M.~Seniuch,
  JHEP {\bf 0502}, 026 (2005)
  doi:10.1088/1126-6708/2005/02/026
  [hep-ph/0412366].

\bibitem{Fromme:2006cm} L.~Fromme, S.~J.~Huber and M.~Seniuch,
  JHEP {\bf 0611}, 038 (2006)
  doi:10.1088/1126-6708/2006/11/038
  [hep-ph/0605242].

\bibitem{Fromme:2006wx} L.~Fromme and S.~J.~Huber,
  JHEP {\bf 0703}, 049 (2007)
  doi:10.1088/1126-6708/2007/03/049
  [hep-ph/0604159].

\bibitem{Cline:2011mm} J.~M.~Cline, K.~Kainulainen and M.~Trott,
  JHEP {\bf 1111}, 089 (2011)
  doi:10.1007/JHEP11(2011)089
  [arXiv:1107.3559 [hep-ph]].

\bibitem{Shu:2013uua} J.~Shu and Y.~Zhang,
  Phys.\ Rev.\ Lett.\  {\bf 111}, no. 9, 091801 (2013)
  doi:10.1103/PhysRevLett.111.091801
  [arXiv:1304.0773 [hep-ph]].

\bibitem{Dorsch:2016nrg} G.~C.~Dorsch, S.~J.~Huber, T.~Konstandin and J.~M.~No,
  JCAP {\bf 1705}, no. 05, 052 (2017)
  doi:10.1088/1475-7516/2017/05/052
  [arXiv:1611.05874 [hep-ph]].

\bibitem{Basler:2017uxn} P.~Basler, M.~Mühlleitner and J.~Wittbrodt,
  JHEP {\bf 1803}, 061 (2018)
  doi:10.1007/JHEP03(2018)061
  [arXiv:1711.04097 [hep-ph]].

\bibitem{Egana-Ugrinovic:2017jib} D.~Egana-Ugrinovic,
  JHEP {\bf 1712}, 064 (2017)
  doi:10.1007/JHEP12(2017)064
  [arXiv:1707.02306 [hep-ph]].

\bibitem{Bruggisser:2017lhc} S.~Bruggisser, T.~Konstandin and G.~Servant,
  JCAP {\bf 1711}, no. 11, 034 (2017)
  doi:10.1088/1475-7516/2017/11/034
  [arXiv:1706.08534 [hep-ph]].
  
\bibitem{deVries:2017ncy} 
  J.~de Vries, M.~Postma, J.~van de Vis and G.~White,
  JHEP {\bf 1801}, 089 (2018)
  doi:10.1007/JHEP01(2018)089
  [arXiv:1710.04061 [hep-ph]].

\bibitem{Huang:2018aja} F.~P.~Huang, Z.~Qian and M.~Zhang,
  Phys.\ Rev.\ D {\bf 98}, no. 1, 015014 (2018)
  doi:10.1103/PhysRevD.98.015014
  [arXiv:1804.06813 [hep-ph]].

\bibitem{Bruggisser:2018mrt} S.~Bruggisser, B.~Von Harling, O.~Matsedonskyi and G.~Servant,
  arXiv:1804.07314 [hep-ph].

\bibitem{Lee:1973iz} T.~D.~Lee,
  Phys.\ Rev.\ D {\bf 8}, 1226 (1973).
  doi:10.1103/PhysRevD.8.1226

\bibitem{Lee:1974jb} T.~D.~Lee,
  Phys.\ Rept.\  {\bf 9}, 143 (1974).
  doi:10.1016/0370-1573(74)90020-9

\bibitem{Fayet:1974fj} P.~Fayet,
  Nucl.\ Phys.\ B {\bf 78}, 14 (1974).
  doi:10.1016/0550-3213(74)90113-8

\bibitem{Glashow:1976nt} S.~L.~Glashow and S.~Weinberg,
  ``Natural Conservation Laws for Neutral Currents,''
  Phys.\ Rev.\ D {\bf 15}, 1958 (1977).
  doi:10.1103/PhysRevD.15.1958

\bibitem{Buras:2010mh} A.~J.~Buras, M.~V.~Carlucci, S.~Gori and G.~Isidori,
  JHEP {\bf 1010}, 009 (2010)
  doi:10.1007/JHEP10(2010)009
  [arXiv:1005.5310 [hep-ph]].
  
  \bibitem{Pich:2009sp} A.~Pich and P.~Tuzon,
  Phys.\ Rev.\ D {\bf 80}, 091702 (2009)
  doi:10.1103/PhysRevD.80.091702
  [arXiv:0908.1554 [hep-ph]].

\bibitem{Gori:2017qwg} S.~Gori, H.~E.~Haber and E.~Santos,
  JHEP {\bf 1706}, 110 (2017)
  doi:10.1007/JHEP06(2017)110
  [arXiv:1703.05873 [hep-ph]].

\bibitem{Craig:2013hca} N.~Craig, J.~Galloway and S.~Thomas,
``Searching for Signs of the Second Higgs Doublet,''
 arXiv:1305.2424 [hep-ph].

\bibitem{Haber:1989xc} H.~E.~Haber and Y.~Nir,
  Nucl.\ Phys.\ B {\bf 335}, 363 (1990).
  doi:10.1016/0550-3213(90)90499-4

\bibitem{Gunion:2002zf} J.~F.~Gunion and H.~E.~Haber,
 ``The CP conserving two Higgs doublet model: The Approach to the decoupling limit,''
 Phys.\ Rev.\ D {\bf 67}, 075019 (2003).

\bibitem{Egana-Ugrinovic:2015vgy} D.~Egana-Ugrinovic and S.~Thomas,
  arXiv:1512.00144 [hep-ph].

\bibitem{Carena:2013ooa} M.~Carena, I.~Low, N.~R.~Shah and C.~E.~M.~Wagner,
  JHEP {\bf 1404}, 015 (2014)
  doi:10.1007/JHEP04(2014)015
  [arXiv:1310.2248 [hep-ph]].

\bibitem{Haber:2018ltt} H.~E.~Haber,
  arXiv:1805.05754 [hep-ph].

\bibitem{Dev:2014yca} P.~S.~Bhupal Dev and A.~Pilaftsis,
  JHEP {\bf 1412}, 024 (2014)
  Erratum: [JHEP {\bf 1511}, 147 (2015)]
  doi:10.1007/JHEP11(2015)147, 10.1007/JHEP12(2014)024
  [arXiv:1408.3405 [hep-ph]].
  
  \bibitem{Weinberg:1990me} S.~Weinberg,
  Phys.\ Rev.\ D {\bf 42}, 860 (1990).
  doi:10.1103/PhysRevD.42.860
  
\bibitem{Ipek:2013iba} 
  S.~Ipek,
  Phys.\ Rev.\ D {\bf 89}, no. 7, 073012 (2014)
  doi:10.1103/PhysRevD.89.073012
  [arXiv:1310.6790 [hep-ph]].

\bibitem{Bian:2014zka} L.~Bian, T.~Liu and J.~Shu,
  Phys.\ Rev.\ Lett.\  {\bf 115}, 021801 (2015)
  doi:10.1103/PhysRevLett.115.021801
  [arXiv:1411.6695 [hep-ph]].

\bibitem{Inoue:2014nva} S.~Inoue, M.~J.~Ramsey-Musolf and Y.~Zhang,
  Phys.\ Rev.\ D {\bf 89}, no. 11, 115023 (2014)
  doi:10.1103/PhysRevD.89.115023
  [arXiv:1403.4257 [hep-ph]].


\bibitem{Arbey:2017gmh} A.~Arbey, F.~Mahmoudi, O.~Stal and T.~Stefaniak,
  Eur.\ Phys.\ J.\ C {\bf 78}, no. 3, 182 (2018)
  doi:10.1140/epjc/s10052-018-5651-1
  [arXiv:1706.07414 [hep-ph]].

\bibitem{Barr:1990vd} S.~M.~Barr and A.~Zee,
  ``Electric Dipole Moment of the Electron and of the Neutron,''
  Phys.\ Rev.\ Lett.\  {\bf 65}, 21 (1990)
  Erratum: [Phys.\ Rev.\ Lett.\  {\bf 65}, 2920 (1990)].
  doi:10.1103/PhysRevLett.65.21

\bibitem{Leigh:1990kf} R.~G.~Leigh, S.~Paban and R.~M.~Xu,
  Nucl.\ Phys.\ B {\bf 352}, 45 (1991).
  doi:10.1016/0550-3213(91)90128-K

\bibitem{Gunion:1990ce} J.~F.~Gunion and R.~Vega,
  Phys.\ Lett.\ B {\bf 251}, 157 (1990).
  doi:10.1016/0370-2693(90)90246-3

\bibitem{Chang:1990sf} D.~Chang, W.~Y.~Keung and T.~C.~Yuan,
  Phys.\ Rev.\ D {\bf 43}, 14 (1991).
  doi:10.1103/PhysRevD.43.R14

\bibitem{Abe:2013qla} T.~Abe, J.~Hisano, T.~Kitahara and K.~Tobioka,
  JHEP {\bf 1401}, 106 (2014)
  Erratum: [JHEP {\bf 1604}, 161 (2016)]
  doi:10.1007/JHEP01(2014)106, 10.1007/JHEP04(2016)161
  [arXiv:1311.4704 [hep-ph]].

\bibitem{Jarlskog:1985ht} C.~Jarlskog,
  Phys.\ Rev.\ Lett.\  {\bf 55}, 1039 (1985).
  doi:10.1103/PhysRevLett.55.1039

\bibitem{Jung:2013hka} M.~Jung and A.~Pich,
  JHEP {\bf 1404}, 076 (2014)
  doi:10.1007/JHEP04(2014)076
  [arXiv:1308.6283 [hep-ph]].


\bibitem{Pospelov:1991zt} M.~E.~Pospelov and I.~B.~Khriplovich,
  Sov.\ J.\ Nucl.\ Phys.\  {\bf 53}, 638 (1991)
  [Yad.\ Fiz.\  {\bf 53}, 1030 (1991)].

\bibitem{Aad:2016nal} G.~Aad {\it et al.} [ATLAS Collaboration],
  Eur.\ Phys.\ J.\ C {\bf 76}, no. 12, 658 (2016)
  doi:10.1140/epjc/s10052-016-4499-5
  [arXiv:1602.04516 [hep-ex]].

\bibitem{Chen:2017esh} X.~Chen [ATLAS and CMS Collaborations],
  arXiv:1703.07675 [hep-ex].

\bibitem{Zanzi:2017msx} D.~Zanzi [ATLAS and CMS Collaborations],
  Nucl.\ Part.\ Phys.\ Proc.\  {\bf 287-288}, 115 (2017)
  doi:10.1016/j.nuclphysbps.2017.03.057
  [arXiv:1703.10259 [hep-ex]].

\bibitem{Harnik:2013aja} R.~Harnik, A.~Martin, T.~Okui, R.~Primulando and F.~Yu,
  Phys.\ Rev.\ D {\bf 88}, no. 7, 076009 (2013)
  doi:10.1103/PhysRevD.88.076009
  [arXiv:1308.1094 [hep-ph]].

\bibitem{Berge:2013jra} S.~Berge, W.~Bernreuther and H.~Spiesberger,
  Phys.\ Lett.\ B {\bf 727}, 488 (2013)
  doi:10.1016/j.physletb.2013.11.006
  [arXiv:1308.2674 [hep-ph]].

\bibitem{Sun:2013yra} Y.~Sun, X.~F.~Wang and D.~N.~Gao,
  Int.\ J.\ Mod.\ Phys.\ A {\bf 29}, 1450086 (2014)
  doi:10.1142/S0217751X14500869
  [arXiv:1309.4171 [hep-ph]].

\bibitem{Anderson:2013afp} I.~Anderson {\it et al.},
  Phys.\ Rev.\ D {\bf 89}, no. 3, 035007 (2014)
  doi:10.1103/PhysRevD.89.035007
  [arXiv:1309.4819 [hep-ph]].

\bibitem{Chen:2013waa} M.~Chen {\it et al.},
  Phys.\ Rev.\ D {\bf 89}, no. 3, 034002 (2014)
  doi:10.1103/PhysRevD.89.034002
  [arXiv:1310.1397 [hep-ph]].

\bibitem{Chen:2015gaa} C.~Y.~Chen, S.~Dawson and Y.~Zhang,
  JHEP {\bf 1506}, 056 (2015)
  doi:10.1007/JHEP06(2015)056
  [arXiv:1503.01114 [hep-ph]].

\bibitem{Askew:2015mda} A.~Askew, P.~Jaiswal, T.~Okui, H.~B.~Prosper and N.~Sato,
  Phys.\ Rev.\ D {\bf 91}, no. 7, 075014 (2015)
  doi:10.1103/PhysRevD.91.075014
  [arXiv:1501.03156 [hep-ph]].

\bibitem{Belyaev:2015xwa} N.~Belyaev, R.~Konoplich, L.~E.~Pedersen and K.~Prokofiev,
  Phys.\ Rev.\ D {\bf 91}, no. 11, 115014 (2015)
  doi:10.1103/PhysRevD.91.115014
  [arXiv:1502.03045 [hep-ph]].

\bibitem{Buckley:2015vsa} M.~R.~Buckley and D.~Goncalves,
  Phys.\ Rev.\ Lett.\  {\bf 116}, no. 9, 091801 (2016)
  doi:10.1103/PhysRevLett.116.091801
  [arXiv:1507.07926 [hep-ph]].

\bibitem{Berge:2015nua} S.~Berge, W.~Bernreuther and S.~Kirchner,
  Phys.\ Rev.\ D {\bf 92}, 096012 (2015)
  doi:10.1103/PhysRevD.92.096012
  [arXiv:1510.03850 [hep-ph]].

\bibitem{Bian:2017jpt} L.~Bian, N.~Chen and Y.~Zhang,
  Phys.\ Rev.\ D {\bf 96}, no. 9, 095008 (2017)
  doi:10.1103/PhysRevD.96.095008
  [arXiv:1706.09425 [hep-ph]].

\bibitem{Chen:2017bff} X.~Chen and Y.~Wu,
  Eur.\ Phys.\ J.\ C {\bf 77}, no. 10, 697 (2017)
  doi:10.1140/epjc/s10052-017-5258-y
  [arXiv:1703.04855 [hep-ph]].

\bibitem{Hagiwara:2016rdv} K.~Hagiwara, K.~Ma and H.~Yokoya,
  JHEP {\bf 1606}, 048 (2016)
  doi:10.1007/JHEP06(2016)048
  [arXiv:1602.00684 [hep-ph]].

\bibitem{Li:2015kxc} G.~Li, H.~R.~Wang and S.~h.~Zhu,
  Phys.\ Rev.\ D {\bf 93}, no. 5, 055038 (2016)
  doi:10.1103/PhysRevD.93.055038
  [arXiv:1506.06453 [hep-ph]].

\bibitem{Brehmer:2017lrt} J.~Brehmer, F.~Kling, T.~Plehn and T.~M.~P.~Tait,
  Phys.\ Rev.\ D {\bf 97}, no. 9, 095017 (2018)
  doi:10.1103/PhysRevD.97.095017
  [arXiv:1712.02350 [hep-ph]].

\bibitem{Barberio:2017ngd} E.~Barberio, B.~Le, E.~Richter-Was, Z.~Was, D.~Zanzi and J.~Zaremba,
  Phys.\ Rev.\ D {\bf 96}, no. 7, 073002 (2017)
  doi:10.1103/PhysRevD.96.073002
  [arXiv:1706.07983 [hep-ph]].

\bibitem{Han:2016bvf} T.~Han, S.~Mukhopadhyay, B.~Mukhopadhyaya and Y.~Wu,
  JHEP {\bf 1705}, 128 (2017)
  doi:10.1007/JHEP05(2017)128
  [arXiv:1612.00413 [hep-ph]].

\bibitem{Bernlochner:2018opw} F.~U.~Bernlochner, C.~Englert, C.~Hays, K.~Lohwasser, H.~Mildner, A.~Pilkington, D.~D.~Price and M.~Spannowsky,
  arXiv:1808.06577 [hep-ph].

\bibitem{Altmannshofer:2015qra} W.~Altmannshofer, J.~Brod and M.~Schmaltz,
  JHEP {\bf 1505}, 125 (2015)
  doi:10.1007/JHEP05(2015)125
  [arXiv:1503.04830 [hep-ph]].
  
\bibitem{Brod:2013cka} 
  J.~Brod, U.~Haisch and J.~Zupan,
  JHEP {\bf 1311}, 180 (2013)
  doi:10.1007/JHEP11(2013)180
  [arXiv:1310.1385 [hep-ph]].
  
\bibitem{Baron:2013eja} 
  J.~Baron {\it et al.} [ACME Collaboration],
  Science {\bf 343}, 269 (2014)
  doi:10.1126/science.1248213
  [arXiv:1310.7534 [physics.atom-ph]].


\end{thebibliography}
\end{document}